\documentclass[prd,aps,nopacs,floats,letterpaper,floatfix,preprint,nofootinbib]{revtex4}

\usepackage{amssymb,amsmath,amsbsy}
\usepackage[dvips]{graphicx}
\usepackage{dcolumn}
\usepackage[usenames]{color}
\usepackage{euscript}
\usepackage{mathrsfs}
\usepackage{enumerate}
\usepackage{subeqnarray}

\newcommand{\be}{\begin{equation}}
\newcommand{\ee}{\end{equation}}
\newcommand{\ba}{\begin{eqnarray}}
\newcommand{\ea}{\end{eqnarray}}
\newcommand{\bel}[1]{\begin{equation}\label{#1}}
\newcommand{\bal}[1]{\begin{eqnarray}\label{#1}}

\newcommand{\nn}{\nonumber}
\newcommand{\rmd}{{\rm d}}

\newcommand{\bee}{\begin{eqnarray}}
\newcommand{\enn}{\end{eqnarray}}
\newcommand{\erf}[1]{(\ref{#1})}

\def\cqg{{\it Class.~Quant.~Grav.}} 

\newlength{\sizeonefig}
\newlength{\sizetwofig}
\begin{document}

\title{Observable properties of orbits in exact bumpy spacetimes}

\author{Jonathan R Gair}
\email{jgair@ast.cam.ac.uk}
\affiliation{Institute of Astronomy, Madingley Road, Cambridge, CB3 0HA, UK}
\author{Chao Li}
\affiliation{Theoretical Astrophysics, California Institute of Technology, 
	Pasadena, California 91125}
\author{Ilya Mandel}
\affiliation{Theoretical Astrophysics, California Institute of Technology, 
	Pasadena, California 91125}

\begin{abstract} 

We explore the properties of test-particle orbits in ``bumpy'' spacetimes ---
stationary, reflection-symmetric, asymptotically flat solutions of Einstein
equations that have a non-Kerr (anomalous) higher-order multipole-moment
structure but can be tuned arbitrarily close to the Kerr metric. Future detectors should observe gravitational waves generated during inspirals of compact objects into supermassive central bodies. If the central body deviates from the Kerr metric, this will manifest itself in the
emitted waves. Here, we explore some of the features of orbits in non-Kerr spacetimes that might lead to observable signatures. As a basis for this analysis, we use a family of exact solutions proposed by Manko \& Novikov which deviate from the Kerr metric in the quadrupole and higher moments, but we also compare our results to other work in the literature. We examine isolating integrals of the orbits and find that the majority of geodesic orbits have an approximate fourth constant of the motion (in addition to the energy, angular momentum and rest mass) and the resulting orbits are tri-periodic to high precision. We also find that this fourth integral can be lost for certain orbits in some oblately deformed Manko-Novikov spacetimes, leading to ergodic motion. However, compact objects will probably not end up on these chaotic orbits in nature. We compute the location of the innermost stable circular orbit (ISCO) and find that the behavior of an orbit in the approach to the ISCO can be qualitatively different depending on whether the location of the ISCO is determined by the onset of an instability in the radial or vertical direction. Finally, we compute periapsis and orbital-plane precessions for nearly circular and nearly equatorial orbits in both the strong and weak field, and discuss weak-field precessions for eccentric equatorial orbits.

\end{abstract}


\date{\today}

\maketitle


\section{Introduction}
\label{sec:introduction}
The space-based gravitational-wave (GW) detector LISA is expected to detect gravitational waves generated during the inspirals of stellar-mass compact objects (white dwarfs, neutron stars or black holes) into supermassive bodies in the centers of galaxies --- extreme-mass-ratio inspirals (EMRIs). LISA could detect gravitational waves from these systems for several years prior to the plunge of the compact object into the central body and hence observe several hundred thousand waveform cycles. Such observations will provide an exquisite probe of the strong gravity region close to supermassive central bodies (see \cite{Pau2007} for a review).  In principle, the emitted gravitational waveform encodes the multipole structure of the spacetime outside the central object \cite{Ryan1995}. One of the hopes for LISA EMRI observations is to extract this spacetime structure from the data and use it to test whether the central objects are indeed Kerr black holes, as we suppose, or something else~\cite{Ryan1995, Hughes2006}. (Intermediate-mass-ratio inspirals detectable by Advanced LIGO may reveal the spacetime structure outside intermediate-mass central bodies with more modest precision~\cite{Brown2007}.) 

For a Kerr black hole, the spacetime is uniquely determined by the mass and angular
momentum of the hole and all higher multipole moments depend on these in a simple way
\be
M_l + {\rm i}S_l = M({\rm i} \chi M)^l .
\label{nohair}
\ee
Here $M_l$ and $S_l$ are the $l$'th mass and current multipole moments of the gravitational field, $M$ is the mass of the black hole and $\chi$ is its dimensionless spin parameter, $\chi \equiv S_1/M^2 \equiv a/M$. As a consequence of relation~\erf{nohair}, if the quadrupole or higher multipole moments of a supermassive body are measured from an EMRI observation and these are inconsistent with the values
predicted by its mass and spin, the body cannot be a Kerr black hole with a vacuum exterior. The ``no-hair'' theorem states that, in pure gravity, any pseudo-stationary, vacuum and asymptotically flat spacetime containing an event horizon and with no closed timelike curves exterior to the horizon must be described by the Kerr metric~\cite{carter71,robinson75}. If the Cosmic Censorship Conjecture is correct, all astrophysical singularities will be enclosed by a horizon. It is therefore most likely that the supermassive central bodies which are observed to inhabit the nuclei of most galaxies are indeed Kerr black holes. However, LISA should be able to test this assumption. Alternatives to Kerr black holes include ``dirty'' Kerr black holes with external masses (e.g., an accretion disk), exotic supermassive stars such as boson stars \cite{boson}, and naked singularities. ``Hairy'' black hole solutions are also allowed when gravity is coupled to other fields, e.g., a Yang-Mills field (these solutions have been shown to be unstable to perturbations~\cite{straumann90}) or a Skyrme field~\cite{droz91} (stability to generic perturbations is an open question). Sufficiently accurate measurements may allow us to distinguish between these possibilities.

In order to prepare us to interpret LISA observations of EMRIs, to identify any deviations from Kerr that are manifest in the waveforms and even to facilitate detection of inspirals into highly non-Kerr spacetimes, we need to understand how these deviations influence the emitted gravitational waveforms. In an extreme-mass-ratio inspiral, the time-scale for the orbital inspiral due to radiation of energy and angular momentum is generally much longer than the orbital time-scale. We can therefore approximate the inspiral as quasi-stationary, by assuming the inspiraling object is always nearly on a geodesic orbit of the spacetime, and evolving the parameters determining this geodesic slowly over the inspiral (this is usually referred to as the ``adiabatic approximation'' in the literature~\cite{hughes05} since the fluxes of energy and angular momentum used to evolve the sequence of geodesics are computed by assuming the object is on an exact geodesic of the spacetime). In this slow-inspiral limit, the emitted waveforms depend sensitively on the properties of the geodesic orbits in the spacetime --- the dominant frequency components in the gravitational waveform at any moment are harmonics of the orbital frequencies of the underlying geodesic. We can thus understand some of the main consequences of deviations from the Kerr metric by examining the effect of such deviations on test particle orbits in the spacetime. By considering a spacetime with an arbitrary set of multipole moments, Ryan demonstrated that, for nearly circular and nearly equatorial orbits, the periapsis and orbital-plane precessions encoded all of the multipole moments at different orders in a weak field expansion~\cite{Ryan1995}. 

A multipole moment decomposition is not very practical, however, since an infinite number of multipoles are required to characterize the Kerr spacetime. For this reason, Collins \& Hughes \cite{CH2004} and
Glampedakis \& Babak \cite{Glamp2005} took a different approach and explored test particle dynamics
in ``bumpy'' spacetimes, which were constructed as first-order perturbations of the Schwarzschild and Kerr spacetimes respectively and therefore could be made arbitrarily close to Schwarzschild/Kerr by dialing a parameter to zero. Collins \& Hughes coined the phrase ``bumpy'' black hole to describe these spacetimes. In their case, the presence of stresses exterior to the black hole meant that the horizon could be preserved in the presence of the black hole deformation without violating the no-hair theorem. In the present case, this name is not strictly applicable since the spacetimes we consider are not black holes at all, but rather naked singularities not enclosed by an event horizon. However, the term ``bumpy'' black hole is still a good one to describe how the spacetime appears to an observer away from the central object. 

One drawback of the perturbative approach is that the perturbation is not necessarily small close to the central body, and so the first-order perturbation theory used to construct the spacetime breaks down. As a result, the perturbative solutions may only be used relatively far from the central object. In this work, we therefore take an alternative approach and consider the properties of orbits and inspirals in a family of spacetimes that are exact solutions of the vacuum field equations of relativity and which include the Kerr and Schwarzschild spacetimes in a certain limit. We use a family of spacetimes that were derived by Manko \& Novikov~\cite{MN1992}.  As exact solutions, the spacetimes are valid everywhere and can thus be used to probe the orbital dynamics in the strong-field as well as the weak-field. The family has an infinite number of free parameters, which can be chosen to make the multipole moments of the spacetime match those of the Kerr spacetime up to a certain order, and then deviate at higher order. In this paper, we choose to make the multipole moments deviate at the mass quadrupole order and higher, by varying a single parameter, although the formalism generalizes to other types of deviation. We use this family of spacetimes as a test-bed for an exploration of various observable consequences of deviations from the Kerr metric, but we compare to previous work in the literature as we proceed.

The main new results of the current work are as follows. By studying the properties of orbits in the strong field of the spacetime, we find that most geodesics in the spacetime appear to have a fourth isolating integral of the motion, in addition to the energy, angular momentum and rest mass that are guaranteed by the stationarity and axisymmetry of the metric. The corresponding orbits are triperiodic to high accuracy. This was not guaranteed, since the separability of the geodesic equations in Kerr and corresponding existence of a fourth integral (the Carter constant) was unusual. Additionally, we find that for some oblate perturbations of the Kerr spacetime, there are regions of the spacetime in which there appears to be no fourth integral, leading to ergodic motion. If observed, ergodicity would be a clear `smoking-gun' for a deviation from Kerr. Ergodic motion has been found in other exact relativistic spacetimes by other authors, although these investigations were not carried out in the context of their observable consequences for EMRI detections. Sota, Suzuki and Maeda~\cite{ssm96} described chaotic motion in the Zipoy-Voorhees-Weyl and Curzon spacetimes; Letelier \& Viera~\cite{LV97} found chaotic motion around a Schwarzschild black hole perturbed by gravitational waves; Gu\'{e}ron
\& Letelier observed chaotic motion in a black hole spacetime with a dipolar halo~\cite{GL01} and in prolate Erez-Rosen bumpy spacetimes~\cite{Gueron2002}; and Dubeibe, Pachon, and Sanabria-Gomez found that some oblate spacetimes which are deformed generalizations of the Tomimatsu-Sato spacetime could also exhibit chaotic
motion~\cite{Dubeibe2007}. The new features of our current results are the presence of potentially ergodic regions for a wider range of magnitudes of the perturbation, and an examination of whether the ergodic regions are astrophysically relevant. We find that, in the context of an EMRI, the ergodic regions exist only very close to the central body and these regions are probably not astrophysically accessible, at least in the Manko-Novikov spacetime family.

We also look at the properties of the last stable orbit for circular equatorial inspirals. The frequency of this orbit will be a gravitational-wave observable, and depends significantly on the magnitude of any deviations from Kerr. For certain choices of the quadrupole perturbation, we find that the last stable orbit is defined by the onset of a vertical instability, rather than the radial instability which characterizes the last stable orbit in Kerr. This is a qualitative observable that could be another `smoking-gun' for a deviation from Kerr.

Finally, we look at the periapsis and orbital-plane precession frequencies. We do this primarily for nearly circular and nearly equatorial orbits, since these can be characterized in a gauge invariant way in terms of the orbital frequency measured by an observer at infinity. Although such precessions were computed by Ryan~\cite{Ryan1995}, his results only apply in the weak-field. We find results that are consistent with Ryan's in the weak-field, but also explore the properties of precessions in the strong-field and find they depend significantly on the nature and location of the last stable orbit. Collins \& Hughes~\cite{CH2004} and Glampedakis \& Babak~\cite{Glamp2005} did explore strong-field precessions, but they did so as a function of spacetime coordinates, rather than as a function of observable quantities which we do here. The perturbative spacetimes are also not totally applicable in the vicinity of the last stable orbit, so our results are more generally applicable. We also briefly discuss  precessions for eccentric equatorial orbits in the weak-field and how this is relevant for LISA observations.

The paper is organized as follows. In Sec.~\ref{spacetimes}, we introduce our chosen family of spacetimes, describe some properties of these solutions and discuss  our approach to computing
geodesics in the spacetimes.  In Sec.~\ref{integrals} we analyze
geodesics in these bumpy spacetimes and use Poincar\'{e} maps to identify the presence of an effective fourth integral of the motion. We show that most orbits are regular and triperiodic, but also demonstrate the onset of ergodic motion in certain oblately deformed spacetimes. In Sec.~\ref{LSOsec} we find the last stable orbit for circular equatorial orbits and discuss its properties. In Sec.~\ref{precession} we report our results on the periapsis precession and orbital-plane precession in these spacetimes. Finally, in Sec.~\ref{future} we summarize our results and discuss further extensions to this work. This paper also includes two appendices, in which we present results demonstrating ergodic motion in Newtonian gravity (Appendix~\ref{newtchaos}) and an expansion of the precessions in the weak-field (Appendix~\ref{wfprec}). Throughout this paper we will use units such that $c=G=1$.

\section{Bumpy black hole spacetimes} 
\label{spacetimes}

In this section, we briefly summarize the Manko-Novikov metric \cite{MN1992}. 
This is the test metric for which we will explore the dynamics of orbits in
Sections~\ref{integrals}--\ref{precession}. The Manko-Novikov metric is an
exact stationary, axisymmetric solution of the vacuum Einstein equations that
allows for deviations away from the Kerr spacetime by a suitable choice of
parameters characterizing the higher-order multipole moments.  The presence of
these deviations destroys the horizon, so this is no longer a black-hole
spacetime.  However, its geometry is very similar to that of a Kerr black hole
with  additional anomalous multipole moments until close to the expected
horizon location.  We choose a subclass of the Manko-Novikov metric,
parametrized by a parameter $\beta$. For $\beta=0$, the metric corresponds to
the usual Kerr metric.  (In the notation of \cite{MN1992}, our parametrization
corresponds to setting $\alpha_2=\beta$ and $\alpha_n=0$ for all $n \neq 2$). 

This subclass of the Manko-Novikov metric can be described by a
Weyl-Papapetrou line element in prolate spheroidal coordinates as 
(cf.~Eq.~(1) of \cite{MN1992}):
\bel{MNmetric}
ds^2 =
    -f(dt - \omega d\phi)^2 + k^2 f^{-1} e^{2\gamma} (x^2 - y^2)
    \left( \frac{dx^2}{x^2-1} + \frac{dy^2}{1-y^2}\right ) +
    k^2 f^{-1}(x^2-1)(1-y^2)d\phi^2,
\ee 
where (cf.~Eqs.~(9, 10, 12, 13 of \cite{MN1992}):
\begin{subequations}\label{MNsupport}
\ba
f &=&e^{2\psi}{A/B}, \\
\omega &=&  2k e^{-2\psi}C A^{-1} - 4 k \alpha (1-\alpha^2)^{-1},  \\
e^{2\gamma} &=&\exp{\left(2\gamma^{'}\right)} A(x^2-1)^{-1}(1-\alpha^2)^{-2},\\
A&=&(x^2-1)(1+ab)^2-(1-y^2)(b-a)^2,\\
B&=&[x+1+(x-1)ab]^2+[(1+y)a+(1-y)b]^2,\\
C&=&(x^2-1)(1+ab)[b-a-y(a+b)]+(1-y^2)(b-a)[1+ab+x(1-ab)],\\
\psi&=&\beta R^{-3} P_2,\\
\gamma^{'}&=&\frac{1}{2}\ln{\frac{x^2-1}{x^2-y^2}}+
	\frac{9\beta^2}{6 R^6} (P_3 P_3 - P_2 P_2)\\
	\nn
&+&
	\beta \sum_{\ell=0}^{2} 
	\left(\frac{x-y+(-1)^{2-\ell}(x+y)}{R^{\ell+1}}P_\ell-2\right),\\
a(x,y)&=&-\alpha\exp\left(-2\beta 
    	\left( -1 + \sum\limits^{2}_{\ell=0}\frac{(x-y)P_\ell}{R^{\ell+1}}
	\right)  \right) \label{eq:axy},\\
b(x,y)
&=&
    \alpha\exp\left(2\beta 
    	\left( 1 + \sum\limits^{2}_{\ell=0}\frac{(-1)^{3-\ell}(x+y)P_\ell}
	{R^{\ell+1}}\right)  \right) \label{eq:bxy},\\
R&\equiv&(x^2+y^2-1)^{1/2},\\
P_n&\equiv&P_n(xy/R) \qquad {\rm where\ } 
	P_n(x)=\frac{1}{2^n n!} \left( \frac{d}{dx}\right)^n (x^2-1)^n.
\ea
\end{subequations}
Here $k$, $\alpha$, and $\beta$ are free parameters which determine the
multipole moments of this spacetime.  The first few multipole moments have the
following values (we correct a typo in Eq.~(14) of \cite{MN1992} following 
\cite{HuaThesis}):
\bel{MnJn}
\begin{array}{r@{\ =\ }l@{\qquad}r l}
M_0& k(1+\alpha^2)/(1-\alpha^2) & S_0= & 0 \\
M_1& 0 & S_1= & -2\alpha k^2(1+\alpha^2)/(1-\alpha^2)^2\\
M_2& -k^3[\beta + 4\alpha^2(1+\alpha^2)(1-\alpha^2)^{-3}] &
S_2 = & 0 \\
M_3& 0
& S_3= & 4\alpha k^4 [\beta + 2\alpha^2
(1+\alpha^2)(1-\alpha^2)^{-3}]/(1-\alpha^2) .
\end{array}
\ee
Therefore, for a given choice of mass $M \equiv M_0$, spin $\chi \equiv
S_1/M^2$ and anomalous (additional to Kerr) dimensionless quadrupole moment $q \equiv -(M_2 - M_2^{\rm Kerr})/M^3$, the three metric parameters are:
\bel{kalphabeta}
\alpha=\frac{-1+\sqrt{1-\chi^2}}{\chi}, \qquad k=M\frac{1-\alpha^2}{1+\alpha^2},
	\qquad \beta=q \frac{M^3}{k^3}.
\ee
A given choice of $M$, $\chi$ and $q$ uniquely defines the metric.  With this definition of $q$, a choice $q>0$ represents an oblate perturbation of the Kerr metric, while $q<0$ represents a prolate perturbation. A spacetime is oblate if it has $M_2 < 0$, e.g., for Kerr $M_2 = -\chi^2 M^3$. When we say a prolate/oblate perturbation we mean a perturbation that makes the spacetime more prolate/oblate relative to Kerr. In particular, for $-\chi^2 < q<0$ the spacetime is still oblate, although it has a prolate perturbation relative to the Kerr metric. We note that taking $q\neq0$ changes all higher moments from their Kerr values, so these solutions deviate not only in the mass quadrupole moment but also in the current octupole moment, the mass hexadecapole moment etc.

To present our results, we find it useful to display them in terms of
cylindrical coordinates $\rho$, $z$ and $\phi$.  These are related to the
prolate spheroidal coordinates $x,\ y$ by \cite{HuaThesis}
\be
\rho = k (x^2 - 1)^{1/2} (1-y^2)^{1/2},\qquad z = k x y,
\ee
and the line element in cylindrical coordinates is
\be
ds^2 =
    -f(dt -\omega d\phi)^2 +
    f^{-1}\left[e^{2\gamma}(dz^2 + d\rho^2) + \rho^2d\phi^2\right].
    \label{genmet}
\ee




\subsection{Spacetime Properties}
\begin{figure}
\centering
\begin{tabular}{ccc}
\includegraphics[keepaspectratio=true,width=2.25in,angle=0]{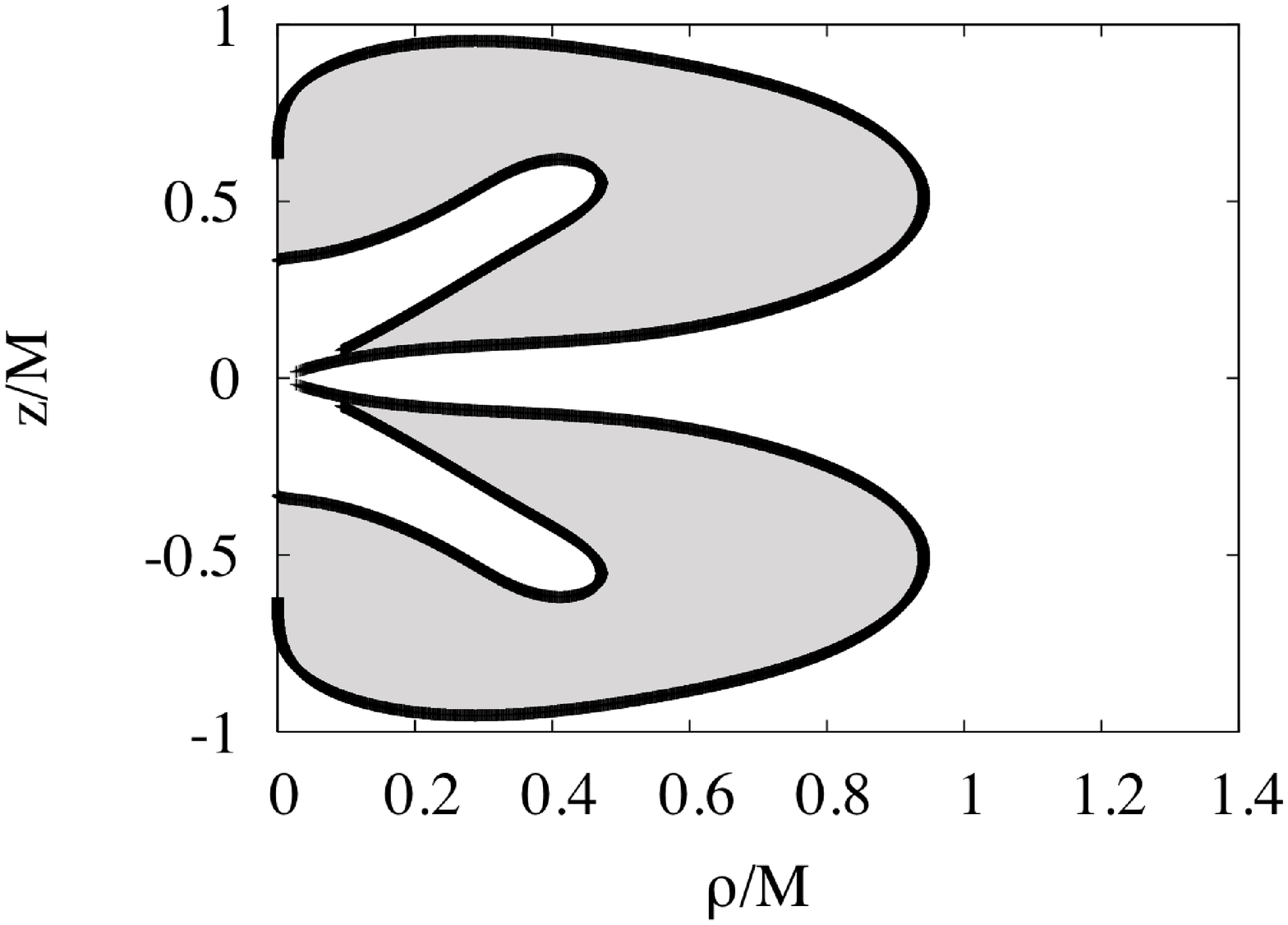} &
\includegraphics[keepaspectratio=true,width=2.25in,angle=0]{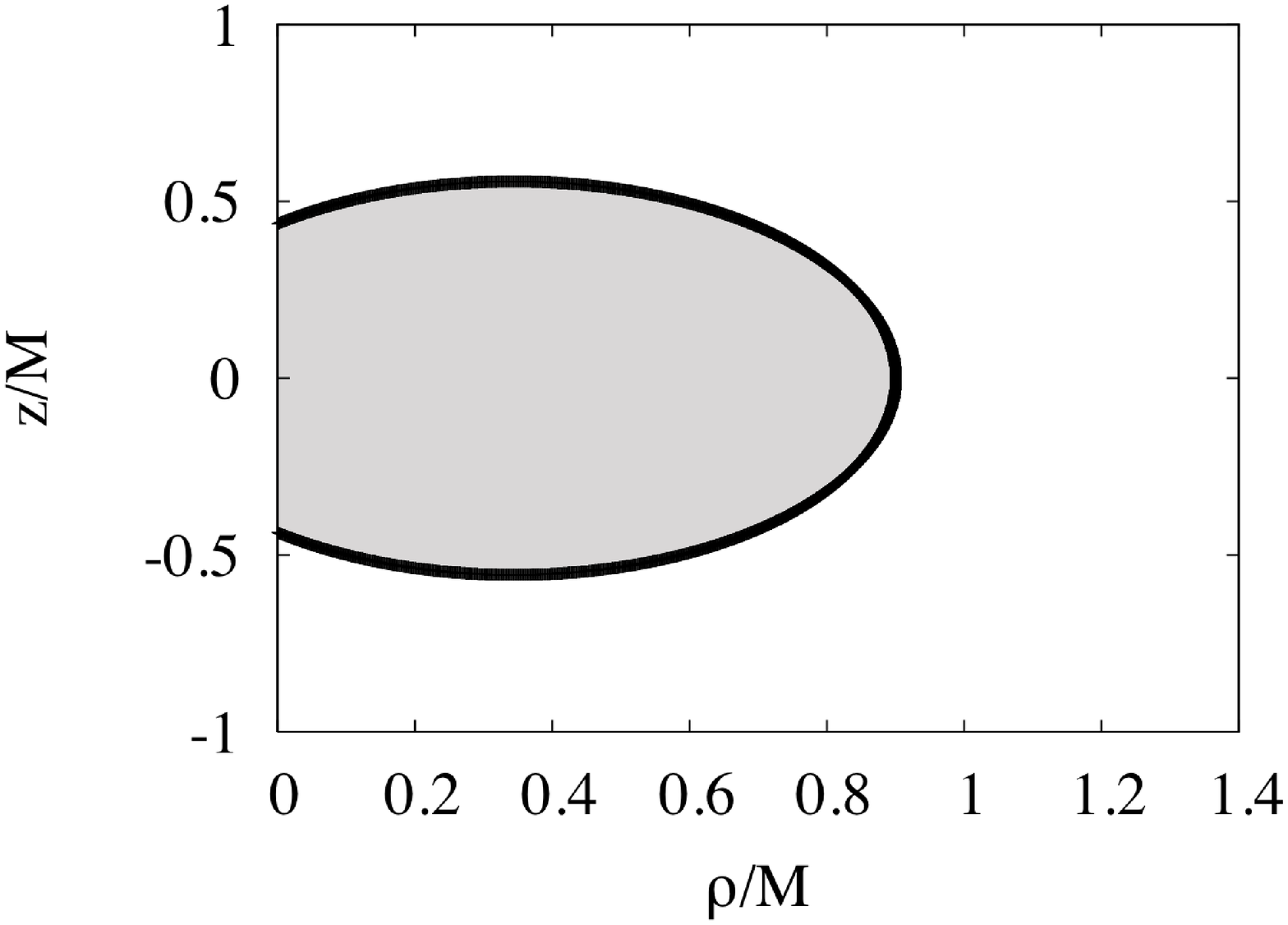} &
\includegraphics[keepaspectratio=true,width=2.25in,angle=0]{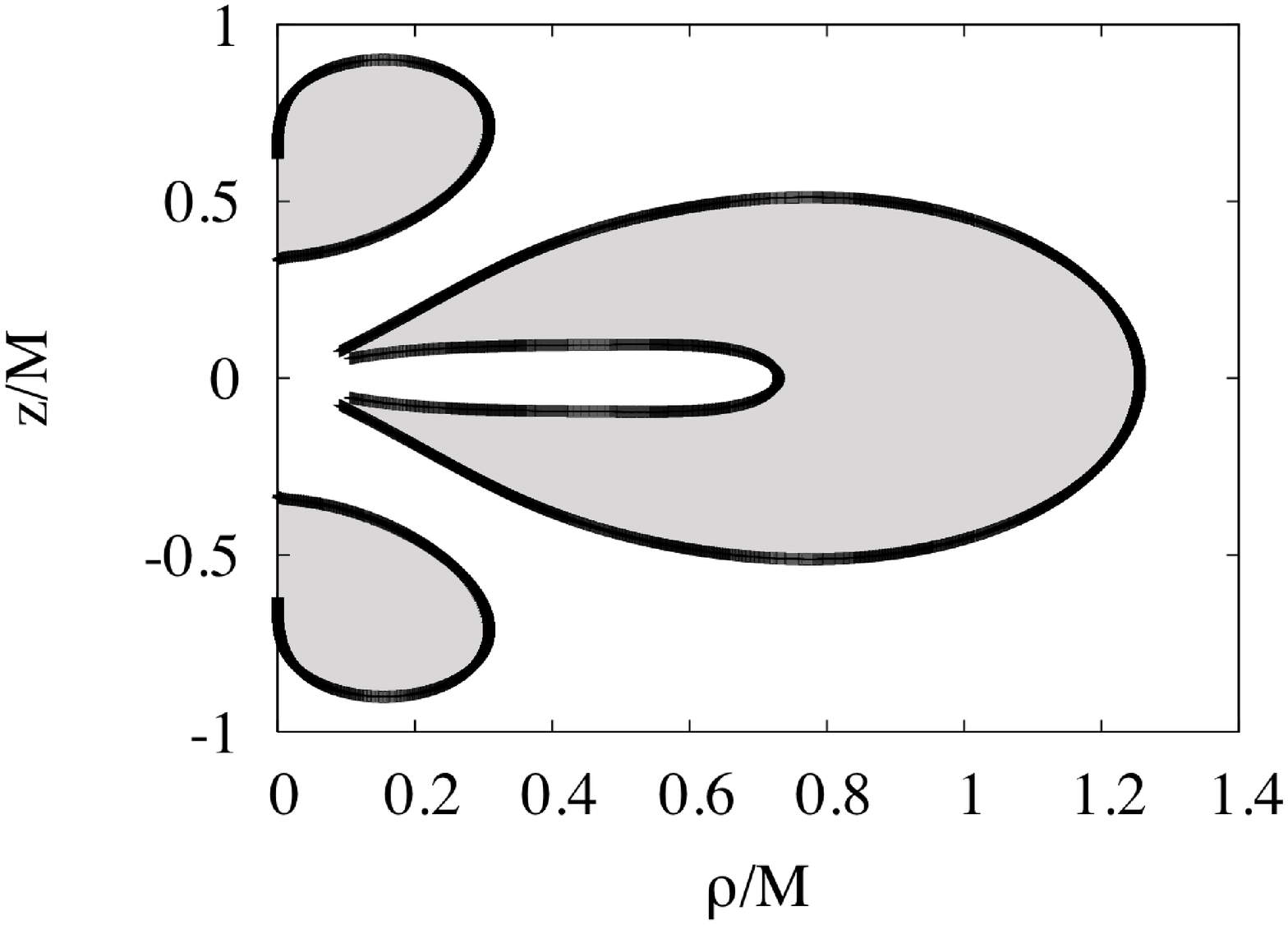} \\
\includegraphics[keepaspectratio=true,width=2.25in,angle=0]{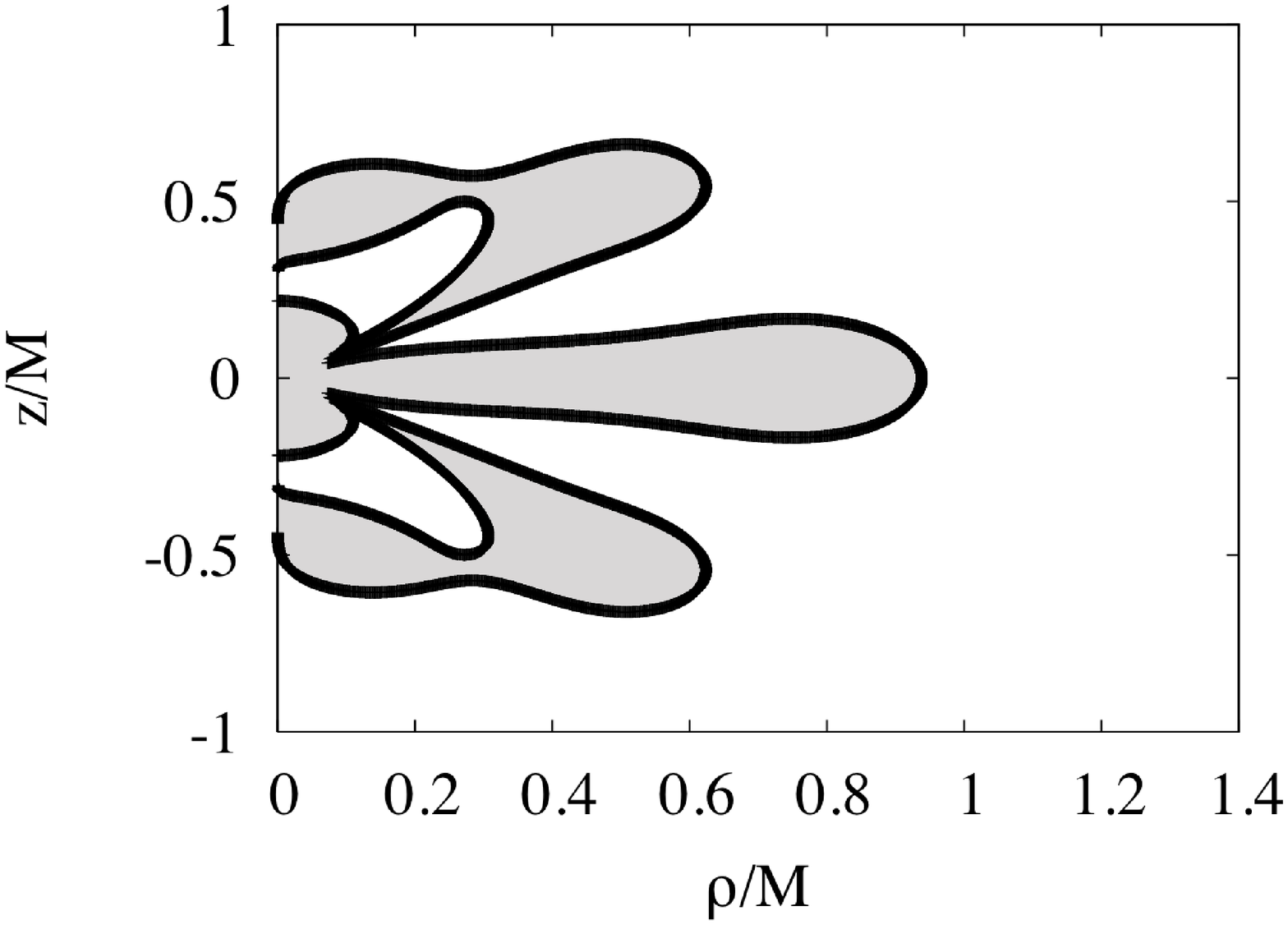} & &
\includegraphics[keepaspectratio=true,width=2.25in,angle=0]{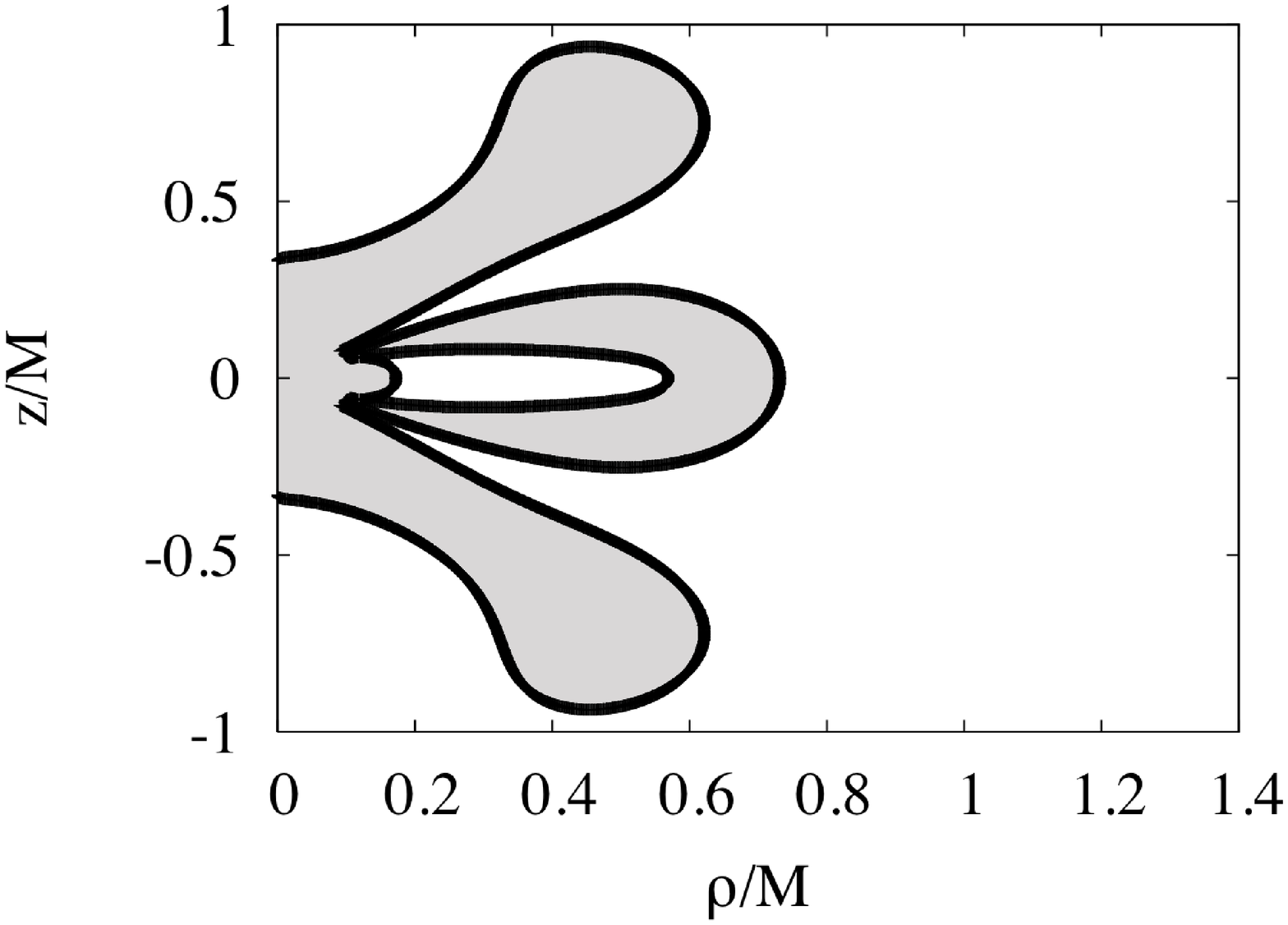} \\
\end{tabular}
\caption{Spacetime structure for $\chi=0.9$. The upper row shows zeros of $g_{tt}$ for $q=-1$ (left column), $q=0$ (middle column) and $q=1$ (right column). This defines the boundary of the ergoregion of the spacetime. The region with $g_{tt} >0$ is shaded. The bottom row shows points where $g_{\phi\phi}$ changes sign for the same values of $q$, and the region where $g_{\phi\phi} <0$ is shaded. This defines the region where closed timelike curves exist. The middle bottom panel is empty since there is no such region in the Kerr spacetime. The shape of the two boundaries is qualitatively the same for other values of $q$ with the same sign, although both regions grow as $|q|$ is increased.}
\label{spacestrucfig}
\end{figure}

The Manko-Novikov spacetimes are vacuum and have the multipolar structure given in Eq.~\erf{MnJn}. As a consequence of the no-hair theorem, the spacetimes must therefore either lack an event horizon or contain closed timelike curves exterior to a horizon. In fact, both of these statements are true. The central singularity is enclosed by a partial horizon at coordinates $\rho =0$, $|z| \leq k$. However, this horizon is broken in the equatorial plane by a circular line singularity at $x=1,\ y=0$ ($\rho=z=0$)~\cite{Brinkprivate}. For $\chi=0$ the spacetime is otherwise regular, but for $\chi \neq 0$, the spacetimes contain both an ergosphere and a region where closed timelike curves exist. The structure of the spacetimes is quite similar to that of the $\delta =2 $ Tomimatsu-Sato spacetime, as described in~\cite{KH03}. The boundary of the ergosphere is determined by the condition $g_{tt} = 0$. Inside this region, timelike observers cannot be at rest. Such a region is entirely physical, and also exists in the Kerr spacetime, where it is of interest since it allows energy extraction via the Penrose process. We show the location of the ergosphere for $\chi=0.9$ and various choices of $q$ in the top panel of Figure~\ref{spacestrucfig}. The shape of the ergosphere is more complicated when $q\neq0$, having a multiple lobed structure. This structure is also qualitatively different depending on the sign of $q$ --- for $q>0$ there are three separate ergoregions, one of which intersects the equatorial plane, one which is entirely above the equatorial plane and one which is entirely below; for $q<0$ there are only two regions, one of which is entirely above the equatorial plane and one of which is entirely below.

For a metric of this type, the region where closed timelike curves (CTCs) exist is determined by the condition $g_{\phi\phi}<0$. In the bottom panel of Figure~\ref{spacestrucfig} we show the portion of the spacetime where CTCs exist for the same choices of $q$ and $\chi=0.9$. Particles orbiting inside the CTC region are moving backward in time. This is not inconsistent with relativity, but CTC zones are sometimes regarded as unphysical. A spacetime with no CTC zone can be constructed by adding an inner boundary in the spacetime, and just using the portion of the Manko-Novikov solution exterior to that boundary.

The CTC zone again has a multiple lobed structure and is different depending on the sign of $q$. We note in particular that for $q < 0$ the ergosphere does not intersect the equatorial plane, although the CTC region does. For $q>0$ both regions intersect the equatorial plane, and the outermost edge of the CTC region is inside the ergoregion.

\subsection{Geodesic Motion}
Geodesic motion in an arbitrary spacetime is described by the second order
equations
\begin{equation}\label{geo}
\frac{d^2 x^{\alpha}}{d \tau^2} =  
	- \Gamma^{\alpha}_{\beta\gamma} \frac{d x^{\beta}}{d\tau} 
	\frac{d x^{\gamma}}{d \tau}. 
\end{equation} 
where the connection coefficients $\Gamma ^{\alpha}_{\beta\gamma}$ are given by
\bel{connection}
\Gamma^{\alpha}_{\beta\gamma}=\frac{1}{2}g^{\alpha\mu} 
	\left(g_{\mu\beta,\gamma}+g_{\mu\gamma,\beta}-g_{\beta\gamma,\mu}\right).
\ee
The spacetimes we are interested in are axisymmetric and time-independent and
the metric correspondingly has two ignorable coordinates --- $t$ and $\phi$.
There are therefore two constants of geodesic motion: the energy $E$ and the
$z$-component of angular momentum $L_z$, which are given by
\bel{genconsts}
E=-g_{tt}\dot{t}-g_{t\phi}\dot{\phi}, \qquad
	L_z=g_{t\phi}\dot{t}+g_{\phi\phi}\dot{\phi},
\end{equation}
where a dot $\dot{}$ denotes the derivative with respect to proper time $\tau$.
Another first integral of the motion can be obtained from conservation of the
rest mass of the orbiting particle:
\bel{masscons}
	-1 = g_{\alpha\beta} \dot{x^\alpha} \dot{x^\beta}.
\ee

\begin{figure}[ht]
\includegraphics[keepaspectratio=true,width=5in]{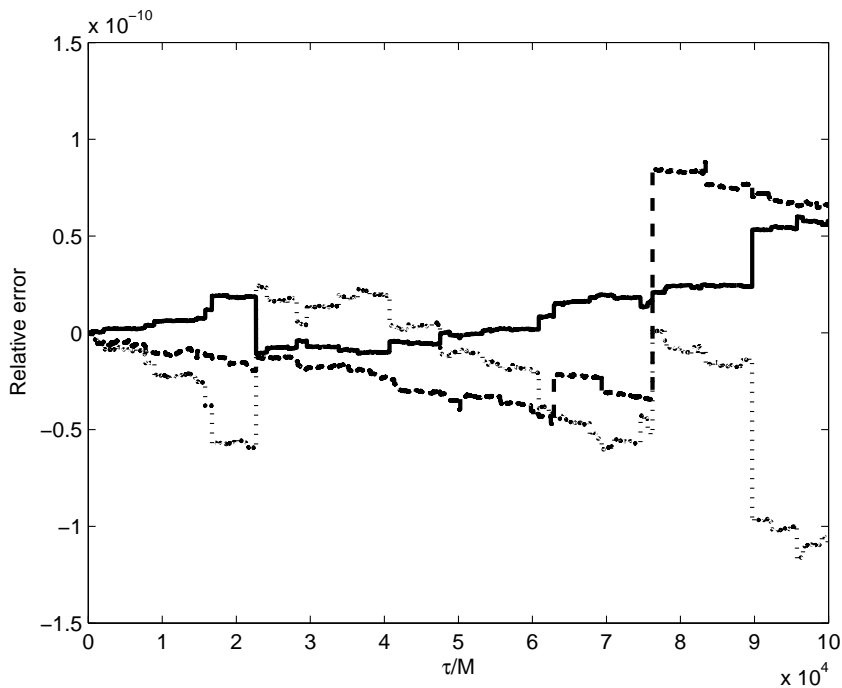}
\caption{The fractional errors in energy $E$ (solid line), 
angular momentum $L_z$ (dashed line), and the quantity 
$g_{\alpha\beta} \dot{x}^\alpha \dot{x}^\beta$ (dotted line) 
accumulated over $1700$ orbits of a geodesic with $E=0.92$ and $L_z=2.5M$
in a spacetime with spin $\chi=0.9$ and anomalous quadrupole moment $q=0.95$.}
\label{fig:errors}
\end{figure}

In practice, we numerically integrate the second-order geodesic
equations~(\ref{geo}) rather than use these first integrals, and we use the
constancy of $E$, $L_z$ and $g_{\alpha\beta} \dot{x}^\alpha \dot{x}^\beta$ as
cross-checks to verify the quality of our numerical results. The results
reported below typically show the conservation of these quantities to a few
parts in $10^{10}$ over the time of integration; see Fig.~\ref{fig:errors}. We
compute the connection coefficients analytically from expressions for the
metric functions $f$, $\omega$ and $\gamma$ defined in Eqs.~(\ref{MNsupport}).
The only difficulty arises at points where a metric component $g_{\mu\nu}$
vanishes and its inverse $g^{\mu\nu}$ diverges. When this occurs, we
analytically factor out the terms that tend to zero to avoid issues in
numerical integration. To perform the numerical integration we write the
coupled system of four second-order ordinary differential equations (\ref{geo})
in first-order form and integrate numerically in C++ via the Bulirsch-Stoer
method.

\begin{figure}[ht]
\includegraphics[keepaspectratio=true,height=5in,angle=-90]{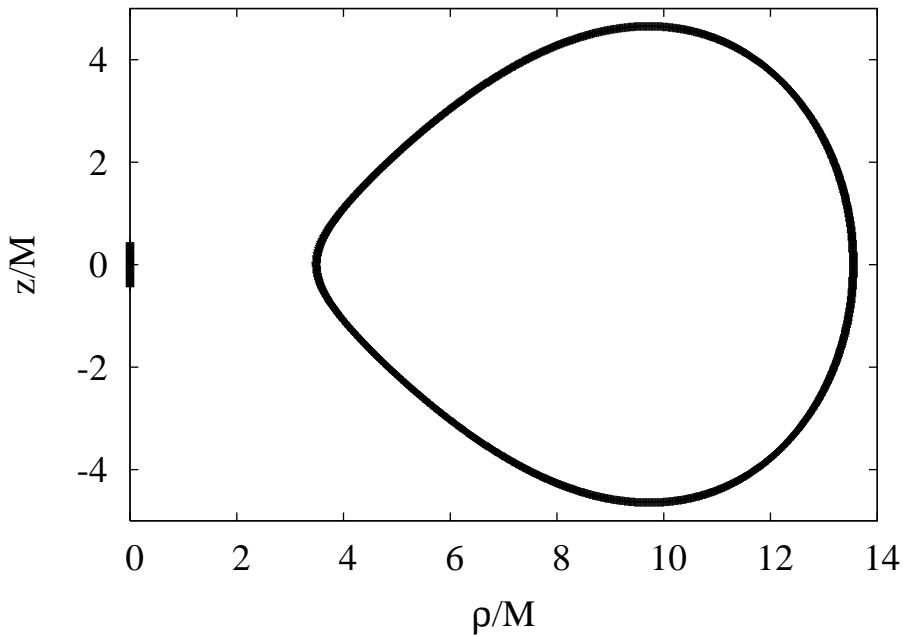}
\caption{Effective potential for geodesic motion around a Kerr black hole, 
with $E=0.95$, $L_z=3 M$ and $\chi=0.9$. The curves indicate zeros of the effective 
potential.  Allowed orbits are found in the small region around $\rho=0$, 
$z=0$ (rising and plunging orbits) or in the region containing $\rho =10$, 
$z=0$ (bound orbits).}
\label{KerrEff}
\end{figure}

Some general properties of geodesic motion can be understood by using the first
integrals~(\ref{genconsts})--(\ref{masscons}). The energy and angular momentum
conservation equations~(\ref{genconsts}) can be used to write $\dot{t}$ and
$\dot{\phi}$ in terms of $E$, $L_z$, $\rho$ and $z$:
\bel{tphidot}
	\dot{t}=\frac{E g_{\phi\phi} + L_z g_{t\phi}}
		{g_{t\phi}^2-g_{tt}g_{\phi\phi}}; \qquad
	\dot{\phi}=\frac{-E g_{t\phi} - L_z g_{tt}}
		{g_{t\phi}^2-g_{tt}g_{\phi\phi}}.
\ee
These expressions can be substituted into Eq.~(\ref{masscons}) to give
\begin{equation}
\frac{{\rm e}^{2\,\gamma(\rho,z)}}{f(\rho,z)}\,
	\left(\dot{\rho}^2 + \dot{z}^2 \right) 
	= \frac{E^2}{f(\rho,z)} - \frac{f(\rho,z)}{\rho^2}\,
		\left[L_z-\omega(\rho,z)\,E \right]^2 - 1 
	\equiv V_{\rm eff}(E, L_z, \rho, z).
\label{Veff}
\end{equation}
The motion in $\rho$ and $z$ may thus be thought of as motion in the effective
potential $V_{\rm eff}$. In particular, since the left hand side of
Eq.~(\ref{Veff}) is strictly positive or zero, motion can only exist in regions
where $V_{\rm eff} \ge 0$. Finding the zeros of the effective potential
therefore allows us to find allowed regions of the motion. As an illustration,
we show the zeros of the effective potential in Figure~\ref{KerrEff} for the
simple case of the Kerr metric with spin parameter $\chi=0.9$, energy $E=0.95$ and
angular momentum $L_z = 3 M$. There are two regions of allowed motion --- one
region at larger radius that corresponds to bound orbits, and another region at
very small radii that corresponds to rising and plunging orbits.

We now turn our attention to the Manko-Novikov spacetime with $q\neq 0$. For spacetimes with $\chi =0$, and for spacetimes with $\chi \neq 0$ and $q<0$ (prolate perturbation of the Kerr metric at large radii), the addition of the perturbation does not fundamentally change the nature of the effective potential -- there are still two bounded regions, one attached to the origin corresponding to rising and
plunging orbits and one at larger radii corresponding to bound orbits. The
shapes of these regions change as $|q|$ is increased and if $|q|$ is increased sufficiently at fixed $E$ and $L_z$ the two regions merge, so
that all allowed orbits can reach the origin. Even after this has occurred,
there appear to be two types of orbit in the single allowed region -- those
that rise and plunge and those that undergo many periods of radial
oscillation.  We don't know if the latter remain non-plunging forever in
principle. In practice, perturbations due to external material or radiation
reaction may cause bound orbits to diffuse onto plunging orbits over time. For fixed $q <0$, the two allowed regions also change shape as the energy and angular momentum are varied. In particular, the plunging region connected to the partial horizon at $\rho = 0$, $|z| \leq k$ develops a multi-lobed structure. For sufficiently large $|q|$ and sufficiently low $E$ and $L_z$, two of these lobes can touch in the equatorial plane. This leads to the existence of circular, equatorial orbits that are unstable to vertical perturbations, which we will encounter again in Section~\ref{LSOsec}.

For $\chi \neq 0$ and  $q > 0$ (oblate perturbation of the Kerr metric at large radii), the behavior is
qualitatively different. For any arbitrarily small $|q|$, an additional allowed
region appears in the effective potential, which is bounded away from $\rho =
0$ and therefore corresponds to bound orbits. For small $|q|$ this new region
is very close to $\rho = 0$. The other two allowed regions still exist, and
merely change shape as the value of $|q|$ is increased. The additional bound region is always outside the region where closed timelike curves (CTCs) exist, and is therefore in the portion of the spacetime that can be regarded as physical. However, in the plane $z=0$ the outermost edge of the CTC region touches the innermost edge of the region of bound motion. This additional region also extends inside the spacetime ergosphere.

\begin{figure}[ht]
\includegraphics[keepaspectratio=true,width=5in,angle=0]{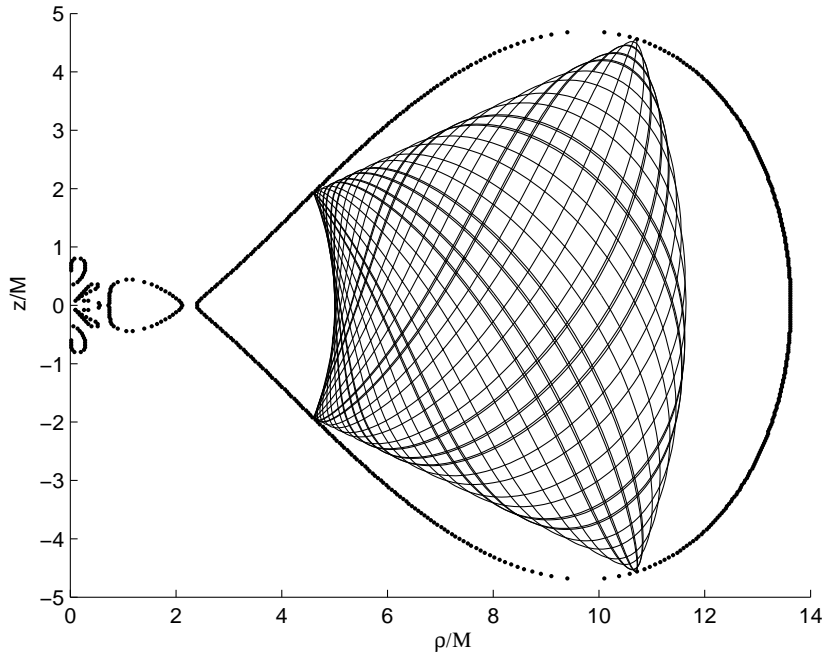}
\caption{Effective potential for geodesic motion around a bumpy black hole with
$\chi=0.9$, $q=0.95$, $E=0.95$, and $L_z=3 M$.  
The thick dotted curves indicate zeros of the effective 
potential.  The trajectory of a typical geodesic in the outer region is shown by a
thin curve.  The regular pattern of self-intersections of the geodesic
projection onto the $\rho-z$ plane indicates (nearly) regular dynamics.}
\label{fig:Veff}
\end{figure}

We consider as an example the case with $\chi=0.9$ and $q=0.95$. The zeros of the
effective potential $V_{\rm eff}$ are plotted in Figure \ref{fig:Veff} for
geodesics with energy $E=0.95$ and angular momentum $L_z=3 M$. 
In this figure there are three distinct allowed regions as described above: (i)
a foliated ``plunging'' region connected to $\rho=0$, where all orbits rapidly
plunge through the horizon (this region also intersects the CTC region); (ii) an inner bound region, which  is located between $\rho/M \approx  0.72$ and $\rho/M \approx 2.12$ for the chosen values of
$E$ and $L_z$;  and (iii) an outer bound region between $\rho/M \approx 2.39$
and $\rho/M \approx 13.6$. We show the trajectory of a typical orbit in the outer region. This has a regular pattern or intersections throughout the $(\rho, z)$ plane, which is characteristic of an orbit with an approximate fourth integral.

If $|q|$ is increased from the value shown in Figure~\ref{fig:Veff}, the two regions of bound motion eventually merge. When this first occurs, the ``neck'' joining the regions is extremely narrow. Geodesics exist which can pass through the neck, but this requires extreme fine tuning. As $|q|$ is
further increased, the neck gradually widens and eventually disappears. At that stage, the single allowed region for bound orbits has a similar shape to the outer region of Figure~\ref{fig:Veff}.

These general properties of the effective potential seem to be common to all spacetimes with $q > 0$ and $\chi \neq 0$. More relevant for the EMRI problem is to fix $q$ and $\chi$ and to vary $E$ and $L_z$. For $E=1$ and sufficiently large $L_z$, there are two regions of allowed motion bounded away from the origin, in addition to the plunging zone connected to $\rho =0$, $|z| \leq k$. The outermost of the allowed regions stretches to infinity and contains parabolic orbits. The inner region of bounded motion is the analogue of the inner bound region described above and lies very close to the central object. If the angular momentum is decreased, while keeping $E=1$, the two non-plunging regions get closer together and eventually merge to leave one allowed region that stretches to infinity. For fixed $E < 1$ the behavior is qualitatively the same, except that for $L_z \gg M$ there is no outer region (there is a maximum allowed angular momentum for bound orbits of a given energy, as in the Kerr spacetime). As $L_z$ is decreased, the outer region for bound motion appears and then eventually merges with the inner region.  Decreasing $L_z$ further eventually causes the bound region to merge with the plunging region. At fixed $L_z$, if there are two distinct non-plunging allowed regions for $E=1$, these regions do not merge as $E$ is decreased, but the outer region eventually disappears (there is a minimum allowed energy for orbits of a given angular momentum, as in the Kerr spacetime). If there is only one non-plunging region for $E=1$, then as $E$ is decreased, this region eventually splits into two allowed regions, and the outer region eventually disappears as $E$ is decreased further. The properties are similar for all $\chi \neq 0$, but decreasing $\chi$ with the other parameters fixed tends to bring the two allowed regions of motion closer to merger with one another.

\section{Isolating integrals}\label{integrals}

The isolating integrals given by the conservation
equations~(\ref{genconsts})--(\ref{masscons}) do not completely describe the
motion, since the motions in $\rho$ and $z$ are coupled. Thus, solution of the
geodesic equations requires use of the second order form of those
equations~(\ref{geo}). However, it was demonstrated by Carter~\cite{Carter68}
that in the Kerr spacetime there is a fourth isolating integral for geodesic
motion, the Carter constant, which arises as a constant of separability of the
Hamilton-Jacobi equation and was later shown to be associated with a Killing
tensor in the spacetime. Carter found the form of all metrics that were both
Schr\"{o}dinger and Hamilton-Jacobi separable. Imposing the further requirement
that the metric be a solution of the vacuum Einstein-Maxwell equations leads
to the Kerr metric as the only spacetime of this form that does not include a
gravomagnetic monopole. Thus, the separability of the equations in Kerr is
somewhat fortuitous and we would not expect that the fourth integral would be
preserved when we add an anomalous quadrupole moment as we do here. As a
consequence, the properties of geodesics might be expected to be somewhat
different, and might even be ergodic.  As mentioned in the introduction, ergodic geodesic motion has been found in other relativistic spacetimes by several other authors~\cite{ssm96,LV97,GL01,Gueron2002,Dubeibe2007}.

A fourth integral of the motion essentially gives another relationship between
$\dot{\rho}^2$ and $\dot{z}^2$. Combining this with the effective potential
equation~(\ref{Veff}) allows us to eliminate $\dot{z}^2$ for instance and hence
obtain an expression for $\dot{\rho}^2$ as a function of $\rho$ and $z$ only.
Similarly we can obtain an expression for $\dot{z}^2$ as a function of $\rho$
and $z$. 

A standard way to examine equations of motion and look for ergodicity is to
plot a Poincar\'{e} map. This involves integrating the equations of motion and
recording the value of $\rho$ and $\dot{\rho}$ every time the orbit crosses a
plane $z=$~constant. From the preceding arguments, if a fourth integral exists,
the value of $\dot{\rho}$ will be a function only of $\rho$ and $z$ (the function
could be multi-valued, depending on the order at which the velocities appear in
the constants of motion). Therefore such a map must show a closed curve.
Similarly, if the Poincar\'{e} map of an orbit shows a closed curve for every
value of $z$, then this defines a relationship between $\dot{\rho}$, $\rho$ and
$z$ which is then an effective fourth integral of the motion. The Poincar\'{e}
analysis thus provides a means to identify whether an effective fourth integral
exists or the motion is apparently ``chaotic''.  In the latter case, the
absence of the integral would  be manifested on the Poincar\'{e} maps as
space-filling trajectories rather than closed curves.

The absence of a full set of isolating integrals does not necessarily
mean that all orbits will exhibit full-blown chaos.  For some initial
conditions, orbits may show obvious signs of ergodicity, while for other
initial conditions in the same spacetime, orbits may appear to behave in an
integrable fashion, suggesting that an approximate additional invariant exists.
Although this behavior may appear surprising at first glance, it is consistent
with the predictions of the KAM theorem and with many known examples of chaotic
behavior.  (The KAM theorem, due to Kolmogorov, Arnold and Moser, states that if
the Hamiltonian of a system with a full set of integrals of motion is
analytically weakly perturbed, then phase-space motion in the perturbed system
will be confined to the neighborhoods of invariant tori in phase space, except
when angle-variable frequencies of the unperturbed system are nearly
commensurate, in which case motion will be chaotic~\cite{KAM}.) 

\begin{figure}[ht]
\includegraphics[keepaspectratio=true,height=5in,angle=-90]{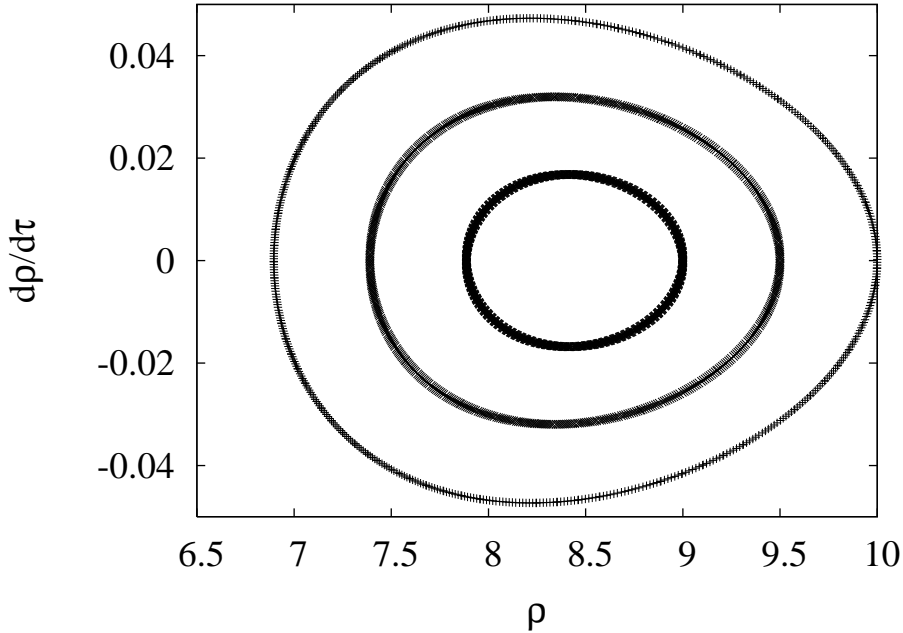}
\caption{Poincar\'{e} map showing $\rmd \rho/\rmd \tau$ vs $\rho$ for 
crossings of the $z=0$ plane for a sequence of orbits in the outer allowed 
region of the Kerr spacetime with $E=0.95$, $L_z=3 M$ and $\chi=0.9$. The closed 
curves indicates the presence of a fourth isolating integral, which we know to 
be the Carter constant.}
\label{KerrPoin}
\end{figure}

As an illustration, we show in Figure~\ref{KerrPoin} the Poincar\'{e} map for
geodesic motion along orbits with three different initial conditions in the
Kerr spacetime with the same $E$, $L_z$ and $\chi$ as Figure~\ref{KerrEff}. The
Poincar\'{e} maps are all closed curves, consistent with the existence of the
fourth isolating integral, the Carter constant. In Appendix~\ref{newtchaos} we present results for motion under gravity in a Newtonian quadrupole-octupole potential and demonstrate the existence of both regular and ergodic orbits. This example serves to put the relativistic results described here in a Newtonian context.

\subsubsection{Poincar\'{e} Maps for the Manko-Novikov spacetimes}
The regularity properties of geodesics appear to be highly correlated with the nature of the effective potential as described in the previous section. For spacetimes with $\chi =0$ and those with $\chi \neq 0$ but $q < 0$, all orbits appear to be regular, i.e., they show closed Poincar\'{e} maps similar to those in Figure~\ref{KerrPoin}. These are the spacetimes in the Manko-Novikov family that have effective potentials which are qualitatively the same as the Kerr case.

\begin{figure}[htb]
\includegraphics[keepaspectratio=true,width=5in,angle=0]{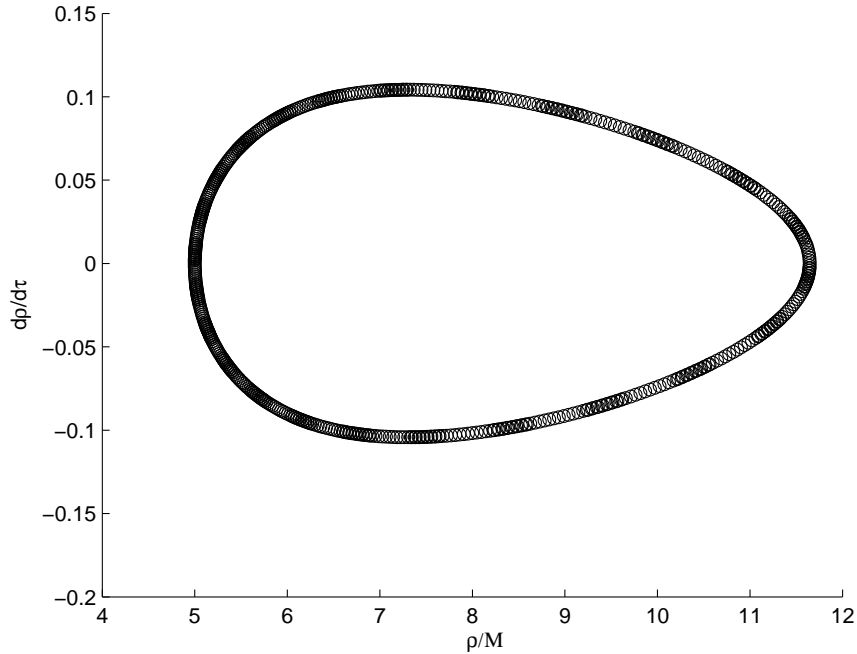}
\caption{Poincare map for a geodesic in the outer region 
	of Fig.~\ref{fig:Veff}.}
\label{fig:Pmapout}
\end{figure}

\begin{figure}[htb]
\includegraphics[keepaspectratio=true,width=5in,angle=0]{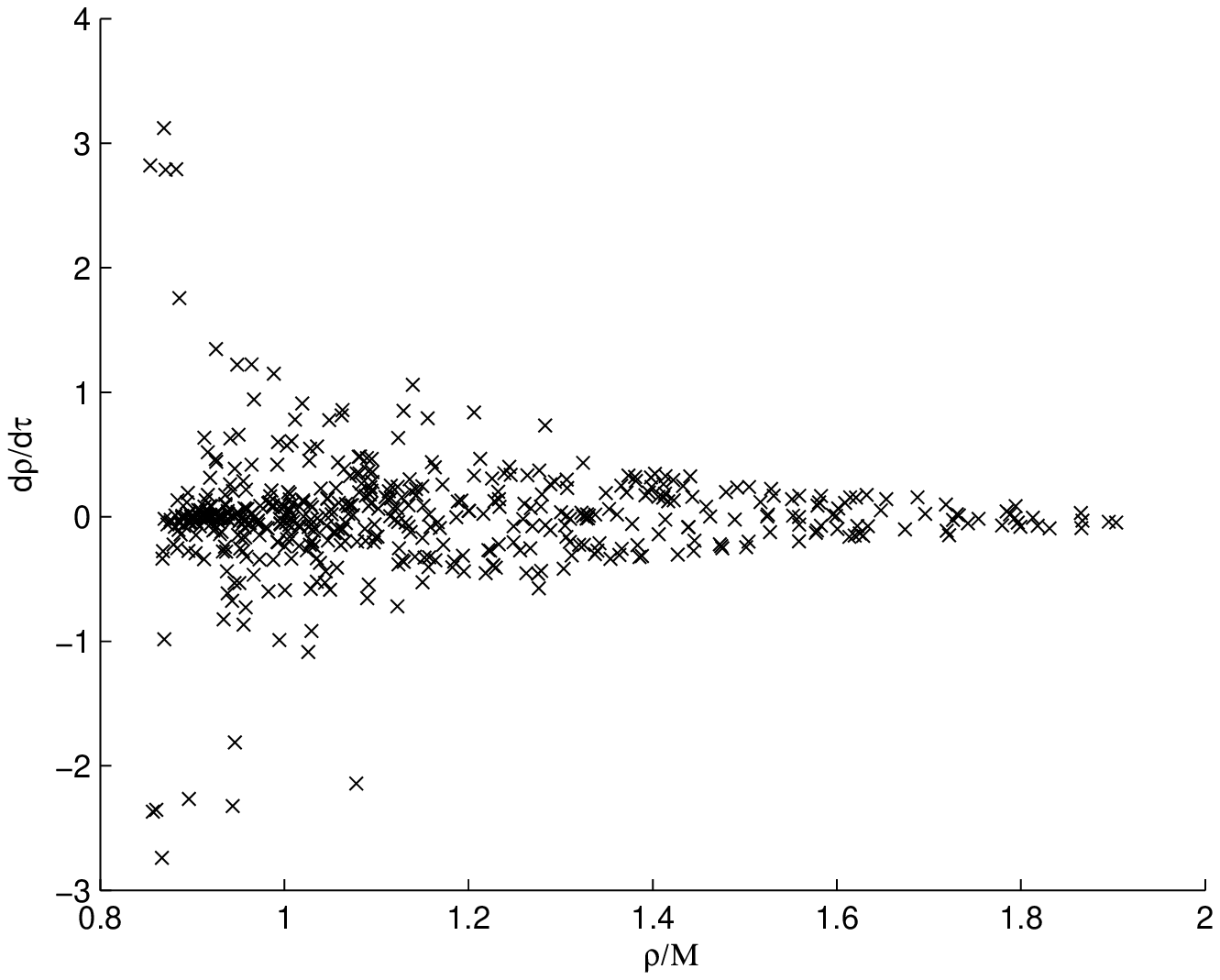}
\caption{Poincare map for a geodesic in the inner region
	of Fig.~\ref{fig:Veff}.}
\label{fig:Pmapin}
\end{figure}

For $q > 0$, the effective potential can have two allowed regions for bound motion. What is striking is
that, whereas orbits in the outer allowed bound region (which corresponds to
the allowed region in the $q=0$ limit) appear to be regular, with closed
Poincar\'{e} maps, those in the inner allowed region appear chaotic. In Figures \ref{fig:Pmapout} and \ref{fig:Pmapin} we show Poincar\'{e} maps for one orbit in each of the outer and inner regions of the effective potential illustrated in Figure~\ref{fig:Veff} ($q=0.95$, $E=0.95$, $L_z = 3M$, $\chi=0.9$). Orbits in the outer region show closed Poincar\'{e} maps, suggesting that the motion
is regular or very nearly so and has an approximate fourth invariant of the motion. This is reinforced by the projection of the orbit onto the $\rho$-$z$ plane, which was shown in Fig.~\ref{fig:Veff}.  The geodesic shows a regular grid pattern, with four possible velocities at each point, corresponding to $\pm |\dot{\rho}|$ and $\pm |\dot{z}|$. If these orbits do not have a true invariant, the regularity of the Poincar\'{e} map suggests that it may still be possible to find an algebraic expression for an approximate constant of the motion.

Orbits in the inner region, by contrast, seem to
fill up all possible points in a subdomain of the allowed parameter space (with $V_{\rm eff} > 0$) and are therefore apparently ergodic in this subdomain.
It seems likely, in view of the KAM theorem, that all orbits in the spacetime
are strictly speaking chaotic, and no true isolating integral exists, but
in the outer region there is a quantity that is nearly invariant
along the orbits~\cite{Brown2007}.   Either the thickness of the
region mapped out by the chaotic motion is small, or the time over which
ergodicity manifests itself is very long. From an observational standpoint,
whether the motion is actually regular or whether only an approximate
invariant exists is irrelevant, since the time-scale over which ergodicity would
manifest itself in the waveform would be much longer than the time during
which the orbiting object moves on an approximate geodesic.

It is unusual, given that chaotic and nearly regular regions are
generally interspersed in most KAM theorem applications \cite{KAM}, that we
find the family of geodesics is divided into two distinct regions such that
geodesics in one region are ergodic while those in the other exhibit nearly
regular orbital dynamics.  We have been unable to find any strongly  ergodic
geodesics in the outer region, or any non-ergodic geodesics in the inner
region. As described in the previous section, adjusting the orbital parameters can cause the two allowed regions to merge. When this first occurs, the two regions are connected by a very narrow neck. The narrowness of the neck means that extreme fine tuning is required to get a geodesic to pass through the
neck. By choosing initial conditions in the neck, and integrating forwards and backwards in
time, we obtained orbits that traversed the neck once and found that the motion was apparently ergodic while in the inner region, but apparently regular in the outer region. This behavior is
consistent with the predictions of the KAM theorem, but observationally the
fact that the orbits in the outer region are technically ergodic does not matter
as long as they appear regular on long time-scales.  We were unable to find an orbit that traversed the neck more than once. Further adjustment of the orbital parameters causes the neck to widen and eventually disappear. At that stage, most of the orbits appear to be regular, but orbits that pass very close to the inner edge of the merged region (i.e., close to the CTC zone) have not been fully investigated.

An alternative explanation of these results \cite{Brink2007} is that the
geodesic equations are numerically unstable in the inner region, and therefore
small numerical round-off errors in the integration routines are driving the orbits away from their true values. Once again, this distinction is not relevant
observationally. An astrophysical system harboring an EMRI will not be
isolated. The gravitational perturbations from distant stars etc.~will serve
the same role in perturbing the orbits as numerical errors might on a computer.
The end result --- that the orbit is apparently ergodic --- is the same.

\subsubsection{Frequency Component Analysis}

\begin{figure}[htb]
\includegraphics[keepaspectratio=true,width=3.5in,angle=0]{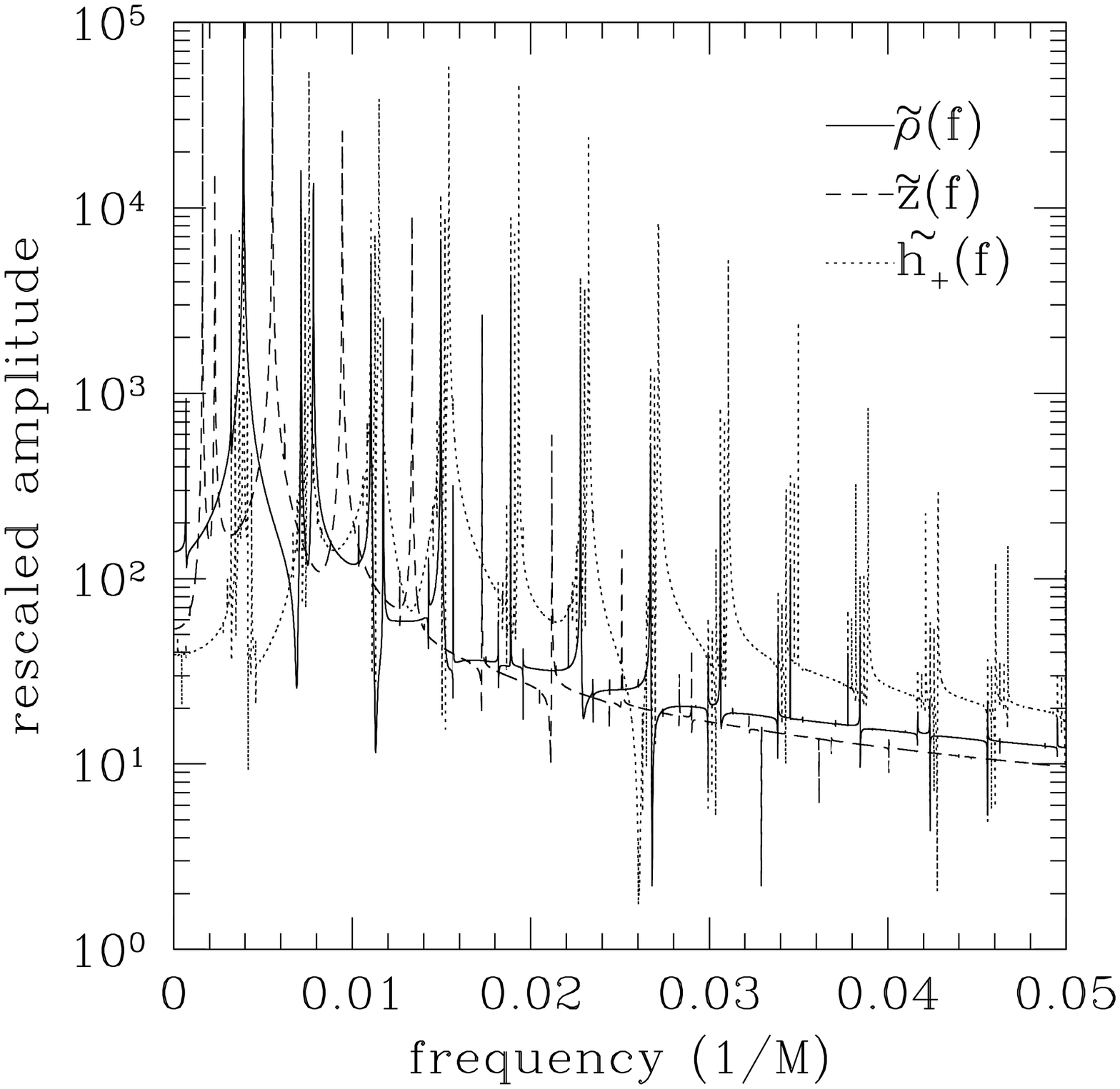}
\caption{Absolute values of the Fourier transforms of $\rho(t)$ (solid line), $z(t)$ (dashed line), 
and the gravitational wave component $h_{+}(t)$ (dotted line) in the frequency domain 
for an orbit in the outer region of
Fig.~\ref{fig:Veff}.  The frequency is displayed in units of $1/M$; the 
amplitude scaling is arbitrary.}
\label{fig:FFTout}
\end{figure}

\begin{figure}[htb]
\includegraphics[keepaspectratio=true,width=3.5in,angle=0]{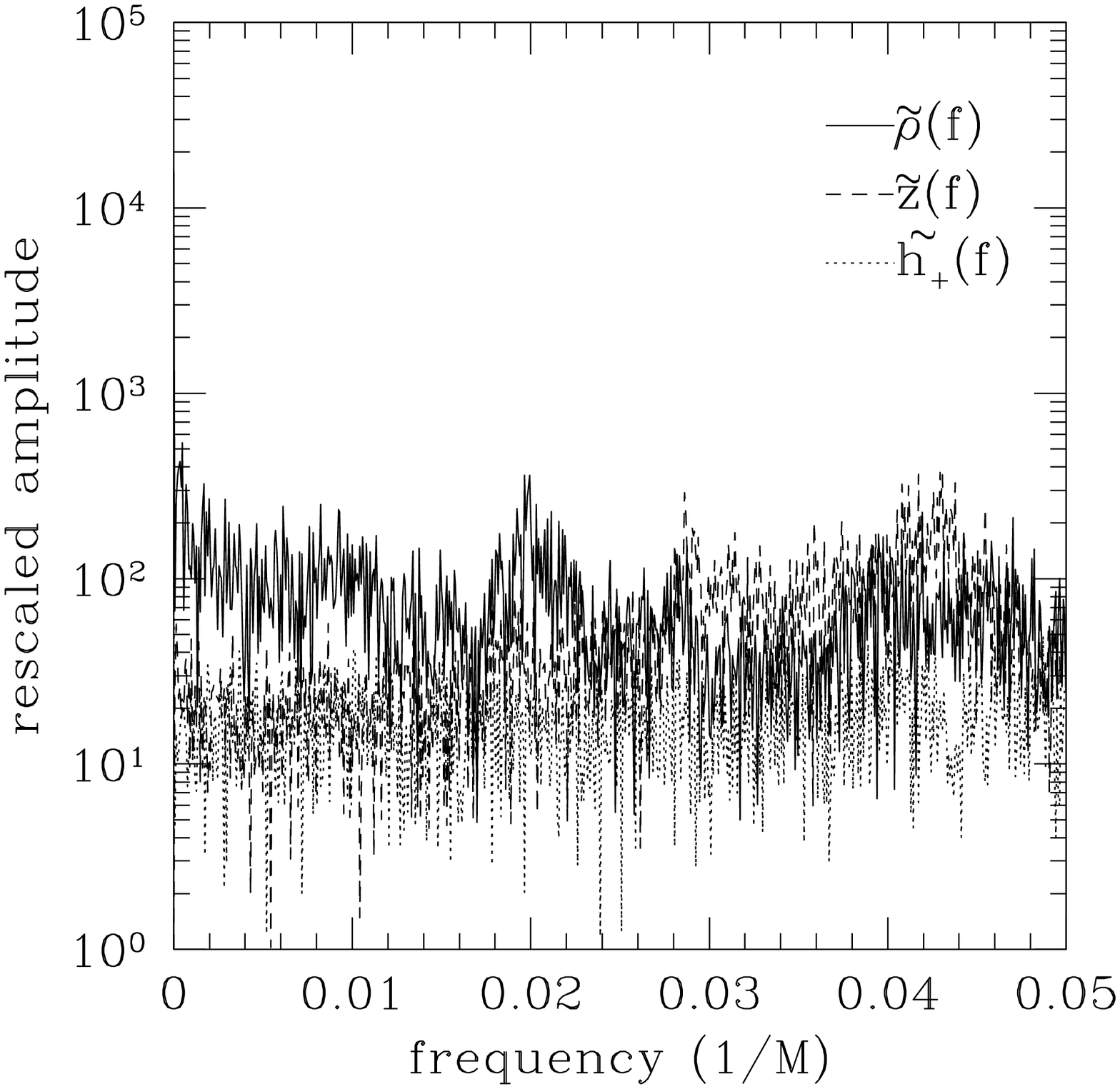}
\caption{Absolute values of the Fourier transforms of $\rho(t)$ (solid line), $z(t)$ (dashed line), 
and the gravitational wave component $h_{+}(t)$ (dotted line) in the frequency domain for an orbit in the 
inner region of
Fig.~\ref{fig:Veff}.  The frequency is displayed in units of $1/M$; the 
amplitude scaling is arbitrary.}
\label{fig:FFTin}
\end{figure}

The above conclusions are supported by a frequency-domain analysis of
the $\rho$ and $z$ motion in the two regions. The absolute values of Fourier
transforms of $\rho(t)$ and $z(t)$ are plotted in Figures~\ref{fig:FFTout} and
\ref{fig:FFTin}. Fig.~\ref{fig:FFTin} shows an absence of clearly identifiable
frequency peaks for geodesics in the inner region, a result consistent with
full-blown chaos.  By contrast, Fig.~\ref{fig:FFTout} shows discrete frequency
peaks in the outer region.  Generally such frequency peaks, corresponding to
harmonics of a few fundamental frequencies, occur in problems with a full set
of isolating integrals.  We find that the frequency components measured for the $\rho$ and
$z$ motion in the outer region can be represented as low order harmonics of two fundamental
frequencies at a high level of precision (1 part in $10^{7}$ for the first $\sim 10$ harmonics). 
This multi-periodicity of the geodesics implies that the gravitational waveforms
will also be multi-periodic. Indeed, we find that an approximate
gravitational waveform, constructed using a semi-relativistic approximation for the 
gravitational-wave emission (as used to construct Kerr EMRI waveforms in~\cite{Babak2007}), is also tri-periodic (the third frequency arises from the $\phi$ motion since the observer is at a fixed sky location).  The absolute value of the
Fourier transform of the $h_+(t)$ component of this gravitational waveform is
also plotted in Fig.~\ref{fig:FFTout} and is clearly multi-periodic.  This
periodicity has important consequences for data analysis and parameter
extraction.  

\subsubsection{Comparison to Other Results}
Our results are consistent with previous work by other authors who have found chaotic geodesic motion in various spacetimes. Generally, chaotic motion only occurs in the strong-field region close to the central object, and for a limited range of geodesic parameters. As an example, Gu\'{e}ron and Letelier~\cite{Gueron2002} found chaos in a prolate Erez-Rosen spacetime, which represented a deformation of a Schwarzschild black hole. They demonstrated that, for a particular value of the energy and angular momentum, when the deformation parameter had a value $k_2=-5$, there was a single allowed region of bounded motion, but for $k_2 = -5.02$ the region split into two separate regions. After the split, orbits in the inner region appeared chaotic while those in the outer region appeared regular. For the merged region, orbits that passed into the inner part also appeared ergodic while those that were purely in the outer part looked regular. This is qualitatively very similar to what we have found in the Manko-Novikov spacetime, although we find chaotic motion only when $\chi\neq0$, while Gu\'{e}ron and Letelier presented examples for both a perturbed non-spinning black hole and a spinning black hole. As a test of our codes, we repeated Gu\'{e}ron and Letelier's calculation and found consistent results. As well as providing another example of chaos for relativistic geodesics, the results here show some new features. In particular, the inner allowed region appears for any $q>0$ and as far as we have been able to ascertain the motion is always ergodic in that region.
This contrasts to the spacetime considered by Gu\'{e}ron and Letelier, in which chaotic motion exists only for a small range of $k_2$ (by the time $k_2$ has increased to $k_2 = -5.1$, the motion is no longer apparently ergodic). Previous authors have also not considered the issue of accessibility of the ergodic region to stars, and we discuss that in the next sub-section.

Sota et al.~\cite{ssm96} discussed what might cause chaos in relativistic geodesic motion, and suggested that it might arise either due to a change in the signs of the eigenvalues of the Weyl tensor, which would lead to ``local instability'' or due to the presence of homoclinic orbits. The Manko-Novikov spacetimes do contain homoclinic orbits, but Sota et al.~\cite{ssm96} found that this only led to chaos in non-reflection symmetric spacetimes, so this explanation probably does not apply here. We have not explored the properties of the eigenspace of the Weyl tensor for these spacetimes, but ``local instability'', could be a  plausible explanation for our results. The CTC region of the Manko-Novikov spacetime might also be causing the ergodicity. The region where ergodic motion occurs touches the CTC region at a single point, so the singular behavior of the metric as the CTC region is approached might explain the observed behavior, either by causing a region of ``local instability'' or through some other mechanism.

We note that in the regime where chaos occurs, the perturbation to the Kerr metric cannot be regarded as purely quadrupolar, but the deviations in the higher multipole moments are also significant. This is similar to the Newtonian result described in Appendix~\ref{newtchaos} since we find chaos in the Newtonian quadrupole-octupole potential but not a pure quadrupole potential. The relativistic results are somewhat different, however, since we find chaos only for $\chi \neq 0$, so for these spacetimes we also need a non-zero current dipole moment to observe chaotic behavior.

\subsubsection{Accessibility of the Ergodic Domain}
While the existence of ergodic motion is mathematically interesting, an important question for EMRIs that has not been addressed so far is whether ergodicity could ever be observed in
nature. In other words, is it possible, during the course of an inspiral, for a
captured object to find itself on an ergodic geodesic?


In typical astrophysical scenarios, the inspiraling compact object will start out far away from the central body with energy close to $1$~\cite{Pau2007}. Unless the angular momentum is very small (which in the Kerr spacetime would represent an object on a plunging orbit), this will correspond to an orbit in the outer region of allowed motion if two regions exist, so the orbit will initially be regular. As the star inspirals, the energy and angular momentum will gradually change and this causes the separation between the outermost point of the inner region of bound motion and the innermost point of the outer region, $\Delta \rho$, to change. For
example, when $E=0.99$ and $L_z=4.33 M$ in a Manko-Novikov spacetime with
$\chi=0.9$, and $q=0.95$, we find that  $\Delta \rho/M \approx 6.4$.  When $E=0.95$ and $L_z=3 M$ in the same spacetime, the separation between regions is only $\Delta \rho \approx 0.27M$.  For
sufficiently small choices of energy and angular momentum (e.g., $E=0.92$ and
$L_z=2.5 M$) only a single region remains. This
suggests that the two regions will come closer together as energy and angular
momentum are radiated away during an inspiral, until they eventually merge.  We conjecture that $\rmd (\Delta \rho)/\rmd t$ is always negative; that is, the
two regions are always merging rather than separating. To test this conjecture,
we must explore the behavior of $\Delta \rho$ along an extreme-mass-ratio
inspiral characterized by slowly evolving $E$ and $L_z$.




To do this, we use an approximate scheme to evolve the energy and angular momentum during
an inspiral.  Our scheme is based on combining exact relativistic expressions
for the evolution of orbital elements with approximate post-Newtonian formulae
for energy and angular-momentum fluxes. This scheme was previously devised to describe EMRIs into
Kerr black holes~\cite{GairGlampedakis2006} and has been shown to give reliable results in that context. For the current calculation, we must augment the fluxes with an additional post-Newtonian term to represent the effect of the
anomalous quadrupole moment $q$ on the evolution of energy and angular
momentum. A Kerr black hole has quadrupole moment $M_2/M^3 = -\chi^2$. It is the quadrupole moment that leads to the lowest order terms in $\chi^2$ in the expressions for the energy and angular momentum radiated during an inspiral. Therefore, to include the excess quadrupole moment, we just change the $\chi^2$ terms in the flux expressions to $\chi^2 + q$, while leaving the lower order terms unchanged (this approach was also used in~\cite{BC2007}). We then numerically find the roots of
the effective potential $V_{\rm eff}=0$ in the equatorial plane at various
times and compute the evolution of $\Delta \rho$ along the inspiral.

\begin{figure}[htb]
\includegraphics[keepaspectratio=true,width=5in,angle=0]{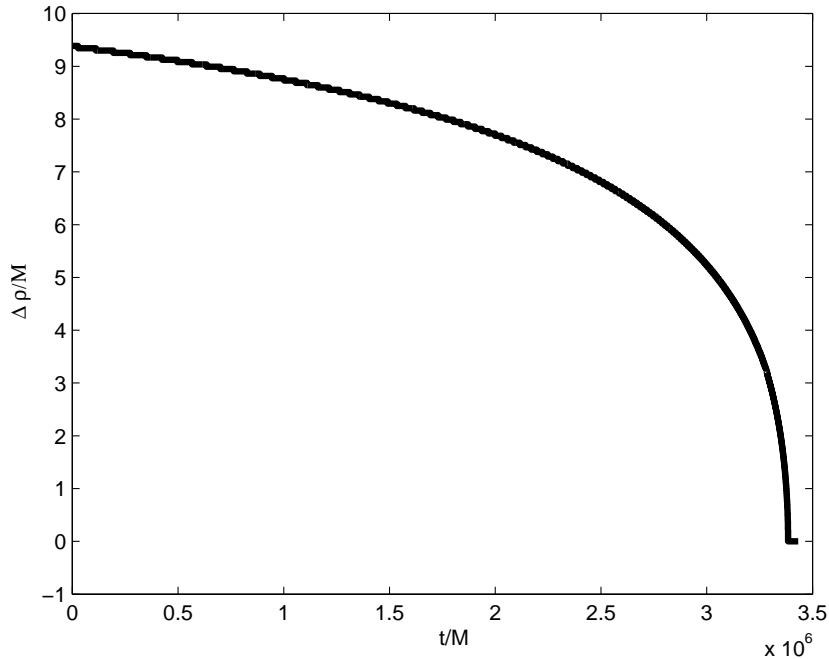}
\caption{The evolution of the separation $\Delta \rho$ between the inner and 
outer bounded regions in the equatorial plane along an inspiral in a
Manko-Novikov spacetime with $\chi=0.9$ and $q=0.95$.  $\Delta \rho=0$ means
that the two regions have merged and there is a single bounded region.}
\label{fig:Deltarho}
\end{figure}

The result of one such computation of $\Delta \rho$ is plotted in
Fig.~\ref{fig:Deltarho}.  That figure corresponds to an inspiral in a spacetime
with $\chi=0.9$, and $q=0.95$.  The inspiral starts out at $\rho=100M$
with an orbital inclination of $60$ degrees and initial eccentricity $e=0.8$ 
(these orbital parameters correspond to $E \approx 0.9982$ and $L_z \approx
5.0852 M$) and proceeds until plunge.  The separation between the inner and outer
bounded regions gradually shrinks, until the two regions merge (on the plot,
this is shown as $\Delta \rho=0$).  Afterward, the bounded regions remain 
joined until eventually merging with the plunging region.

We have found the same qualitative behavior described above for a wide range of parameter choices.  Therefore, in all these cases, our conjecture is true --- the inspiraling object can never find itself in the isolated inner region where all orbits appear to be ergodic.  We
should point out, however, that we have carried out this numerical
investigation only for a range of specific choices of $\chi$, $q$, and initial
orbital parameters, and have used an approximation to the energy and angular momentum radiated during an inspiral. This is therefore not a definitive proof that chaotic motion can
never be observed in the course of an inspiral in the Manko-Novikov spacetime. 

Assuming this evolution really is typical, there are two important consequences. Firstly, an inspiraling object can never end up in the inner of two allowed regions of bound motion, where ergodic motion is prevalent. Secondly, inspirals
always start out in a phase where the motion is regular. This is very
important, since it will allow the systems to be detected in this early inspiral stage by gravitational-wave detectors using matched filtering or a time-frequency analysis. The inspiraling object will eventually end up in the merged region formed after the two regions of bounded motion converge. Both ergodic and regular geodesics exist in that region, so in principle the particle could find itself on an ergodic orbit. However, most orbits in the merged region appear to be regular so it would require fine tuning to put the object onto such a geodesic (e.g., the ``neck traversing'' geodesics discussed earlier). It thus seems unlikely that this would occur in practice.

Although these results apply only to the Manko-Novikov family of spacetimes, the conclusions are consistent with other examples of chaotic geodesics in relativity. For instance, in the prolate Erez-Rosen spacetime considered in~\cite{Gueron2002}, if an object had arrived in the region where ergodic motion is observed during the course of an inspiral, its orbital energy and angular momentum would have been larger earlier in the inspiral. However, if either the energy or angular momentum is increased from the values that give ergodic motion, the effective potential changes so that it has only one allowed region, which includes ``escape zones'' connected to the central singularity. All geodesics in such a zone plunge into the central object in a short time so an astrophysical inspiral could not persist through that zone. We deduce that for that spacetime as well the ergodic region is inaccessible to objects captured at large distances.

If there was some other mechanism that could put an inspiraling object onto an ergodic geodesic, there is the question of how the ergodicity could be identified in practice. Detection of EMRIs will rely on matched filtering or possibly time-frequency techniques~\cite{Pau2007}. In either case, it will probably not be possible to identify the gravitational radiation as being emitted from an ergodic orbit, but only that radiation from a regular orbit has ceased. It is clear from Figure~\ref{fig:FFTin} that during an ergodic phase, the emitted power is spread among many harmonics, which will consequently not be individually resolvable. This radiation will increase the broadband power in our detector, whereas if the orbit had plunged the radiated power would rapidly die away. However, the energy released during a typical EMRI is comparatively low, so it is unlikely that we could identify the presence of such broadband power over the instrumental noise. Therefore, the chances are that we will not be able to distinguish observationally between an inspiral that ``ends'' at a transition into an ergodic phase and one which ends by plunging into a black hole. 

One potentially observable signature of ergodicity would be an inspiral that turned ``off'' and ``on'' as it progressed through ergodic phases interspersed with regular phases. This would occur if the object could move into and out of the inner ergodic region during an inspiral, but the preceding analysis indicates that this shouldn't happen. An object on a ``neck-traversing'' geodesic would also show this behavior. However, the periods where the orbit is ergodic serve to randomize the
phase of the orbit in the regular periods. A signal of this type would only be observable if each apparently regular phase could be individually resolved with enough signal-to-noise
ratio. This would require a very narrow ``neck'' in order to trap the orbit for many cycles in the regular zone. However, fine tuning of the energy and angular momentum is necessary to make the neck very narrow, so if an object was on such an orbit, the neck would be widening rapidly as energy and angular momentum were radiated away. In practice, it is doubtful that sufficient signal-to-noise would accumulate to allow a detection to be made before the neck widened too much.

We conclude that, for astrophysically relevant inspirals in the Manko-Novikov spacetime family, an object would probably not end up on an ergodic geodesic. If some other mechanism conspired to put an object on such an orbit, it is unlikely that we would be able to identify this in gravitational-wave observations. If these findings carry over to a more generic class of spacetimes, then chaotic
motion is merely a mathematical curiosity which is unlikely to manifest itself practically or be important for gravitational-wave data analysis considerations.

\section{Last Stable Orbit}
\label{LSOsec}
During an inspiral into a Kerr black hole, an EMRI will evolve quasi-stationarily through a sequence of near-geodesic orbits as orbital energy and angular momentum are radiated away. There is a minimum energy (which is dependent on angular momentum) for which bound orbits exist. When the inspiral reaches that separatrix, the object will rapidly plunge into the central body. The gravitational radiation emission undergoes a transition at this point, and so the frequency of this last stable orbit is in principle another quantity that is observable from the detected gravitational waves. For a Kerr inspiral, the `transition' is a rapid die-off in the gravitational-wave emission as the particle plunges into the black hole. If the central object is not a black-hole, the radiation may persist for longer after the last stable orbit is passed~\cite{boson}, but there will still be a significant qualitative change in the emitted radiation as the orbit changes suddenly at that point. We focus on the innermost stable circular equatorial orbit in this analysis, since this is well defined in these spacetimes.

\subsection{Circular equatorial orbits}

The geodesic equations for an arbitrary spacetime~(\ref{geo}) may be written 
in the alternative form
\begin{equation}
\frac{\rmd}{\rmd \tau} \left( g_{\mu\alpha} 
	\frac{\rmd x^\alpha}{\rmd \tau}\right) = \frac{1}{2} 
	\partial_\mu g_{\nu\sigma}\,\frac{\rmd x^{\nu}}{\rmd \tau}
	\frac{\rmd x^{\sigma}}{\rmd \tau}.
\label{geoalt}
\end{equation}
For a circular, equatorial orbit in an axi- and reflection-symmetric spacetime
of the form (\ref{genmet}), $\rmd\rho/\rmd\tau=\rmd z/\rmd\tau=
\rmd^2\rho/\rmd\tau^2=0$; hence the $\rho$-component of the geodesic equation (\ref{geoalt}) 
gives
\begin{equation} 
	\partial_\rho g_{\phi\phi} \dot{\phi}^2+2 \partial_\rho
	g_{t\phi} \dot{t} \dot{\phi}+\partial_\rho g_{tt} 
	\dot{t}^2=0,
\end{equation} 
in which a dot denotes $\rmd/\rmd\tau$ as before. We can thus express the
azimuthal frequency as observed at infinity $\Omega_{\phi} \equiv
\dot{\phi}/\dot{t}$ in the form
\bel{Omegaphi}
\Omega_{\phi} = \frac{-\partial_\rho g_{t\phi} \pm 
	\sqrt{(\partial_\rho g_{t\phi})^2-
		\partial_\rho g_{tt}\partial_\rho g_{\phi\phi}}}
		{\partial_\rho g_{\phi\phi}},
\ee
where the $+/-$ signs are for prograde and retrograde orbits respectively. In
the equatorial plane, the right-hand side is a function of the  spacetime
parameters and $\rho$ only, so given a particular choice of azimuthal
frequency  $\Omega_\phi$, Eq.~(\ref{Omegaphi}) can be inverted to determine
the value of $\rho$ such that a circular orbit at that $\rho$ has frequency
$\Omega_\phi$.
 
Equation (\ref{masscons}) provides another relation
between $\dot{t}$ and $\dot{\phi}$, from which we can deduce
\be
\dot{t} = \left(-g_{tt} - 2\Omega_{\phi}g_{t\phi}-
	\Omega_{\phi}^2\,g_{\phi\phi} \right)^{-\frac{1}{2}},
\ee
and then the energy and angular momentum equations (\ref{genconsts}) give us 
$E$ and $L_z$ as a function of $\rho$ for circular equatorial orbits.

\subsection{Innermost Stable Circular Orbit}
The location of the innermost stable circular orbit (ISCO) in the equatorial plane can be found using the effective potential~(\ref{Veff}). Circular equatorial orbits are located at simultaneous zeros and turning points of $V_{\rm eff}$, where $V_{\rm eff} = \partial V_{\rm eff}/\partial \rho = \partial V_{\rm eff}/\partial z = 0$. As we will see in Section~\ref{precession} the second derivatives of $V_{\rm eff}$ determine the frequencies of small oscillations about the circular orbit. For the circular orbit to be stable, we need the orbit to sit at a local maximum of $V_{\rm eff}$, i.e., we require $\partial^2 V_{\rm eff}/\partial \rho^2$ and $\partial^2 V_{\rm eff}/\partial z^2$ to be negative. In the following we will use $\tilde{V}_{\rho\rho}(\rho)$ ($\tilde{V}_{zz}(\rho)$) to denote the value of $\partial^2 V_{\rm eff}/\partial \rho^2$ ($\partial^2 V_{\rm eff}/\partial z^2$) evaluated for the circular equatorial orbit at radius $\rho$. For the Kerr spacetime, $\tilde{V}_{zz}(\rho) < 0$ at all radii, but $\tilde{V}_{\rho\rho}(\rho)$ has a single root at a critical radius $\rho_{\rm ISCO}$. This tells us that the orbit becomes radially unstable at that point, which defines the ISCO. For $\chi = 0$, $\rho_{\rm ISCO} \approx 4.90M$, while for $\chi=0.9$, $\rho_{\rm ISCO} \approx 1.25M$ for prograde orbits and $\rho_{\rm ISCO} \approx 7.705M$ for retrograde orbits. Note that $\rho$ is a cylindrical Weyl coordinate, which is why these results differ from the familiar black-hole ISCO radii, which are normally quoted in Boyer-Lindquist coordinates.

For the Manko-Novikov solutions with $\chi = 0$, the shape of the functions $\tilde{V}_{\rho\rho}(\rho)$ and $\tilde{V}_{zz}(\rho)$ does not change significantly as q is increased with $q>0$: $\tilde{V}_{zz}(\rho)<0$ everywhere and $\tilde{V}_{\rho\rho}(\rho) = 0$ has a single solution that defines the ISCO. However, as $|q|$ is increased with $q<0$, there is a transition in behavior at $q \approx -0.163$. For $q\lesssim -0.163$, the function $\tilde{V}_{\rho\rho}(\rho)$ has two zero-crossings. Thus, in addition to the radially-stable circular orbits at large radii, we find additional such orbits exist very close to the central singularity. If $|q|$ is increased still further, the two roots converge at $q \approx -0.654$ and for $q \lesssim -0.654$ radially stable orbits exist at all values of $\rho$. However, at the point where the second branch of the radial roots appears, there is also a transition in the shape of $\tilde{V}_{zz}(\rho)$, so that there are now orbits which are vertically unstable. For $q\lesssim -0.163$, the ISCO is defined by this vertical instability, rather than the radial instability characteristic of the Kerr spacetime, and Manko-Novikov spacetimes with $q>0$. In the range $-0.654 \lesssim q \lesssim -0.163$, there are two regimes where stable circular orbits exist --- an outer zone with $\rho > \rho_{\rm ISCO}$, and an inner zone with $\tilde{\rho}_{\rm ISCO} < \rho < \rho_{\rm OSCO}$ (we use ``OSCO'' to indicate ``outermost stable circular orbit'' and $\tilde{\rho}_{\rm ISCO}$ to denote the ISCO for the inner set of circular orbits). The energy and angular momentum of an orbit at the ``OSCO'' are greater than the energy and angular momentum at the ISCO of the outer zone, $\rho_{\rm ISCO}$. Thus, an object inspiraling from large distances on a circular equatorial orbit will reach $\rho_{\rm ISCO}$ and plunge into the central body, rather than finding itself in the inner range of circular orbits. Compact objects could only find themselves in the inner range if they came in on an eccentric/inclined orbit and then radiated away energy and angular momentum in exactly the right proportions. It is therefore unlikely that this inner zone would be populated in practice. However, any object on a circular equatorial orbit in this inner zone would reach $\tilde{\rho}_{\rm ISCO}$ and then plunge into the central body.

\begin{figure}
\centering
\begin{tabular}{cc}
\includegraphics[keepaspectratio=true,height=3.5in,angle=-90]{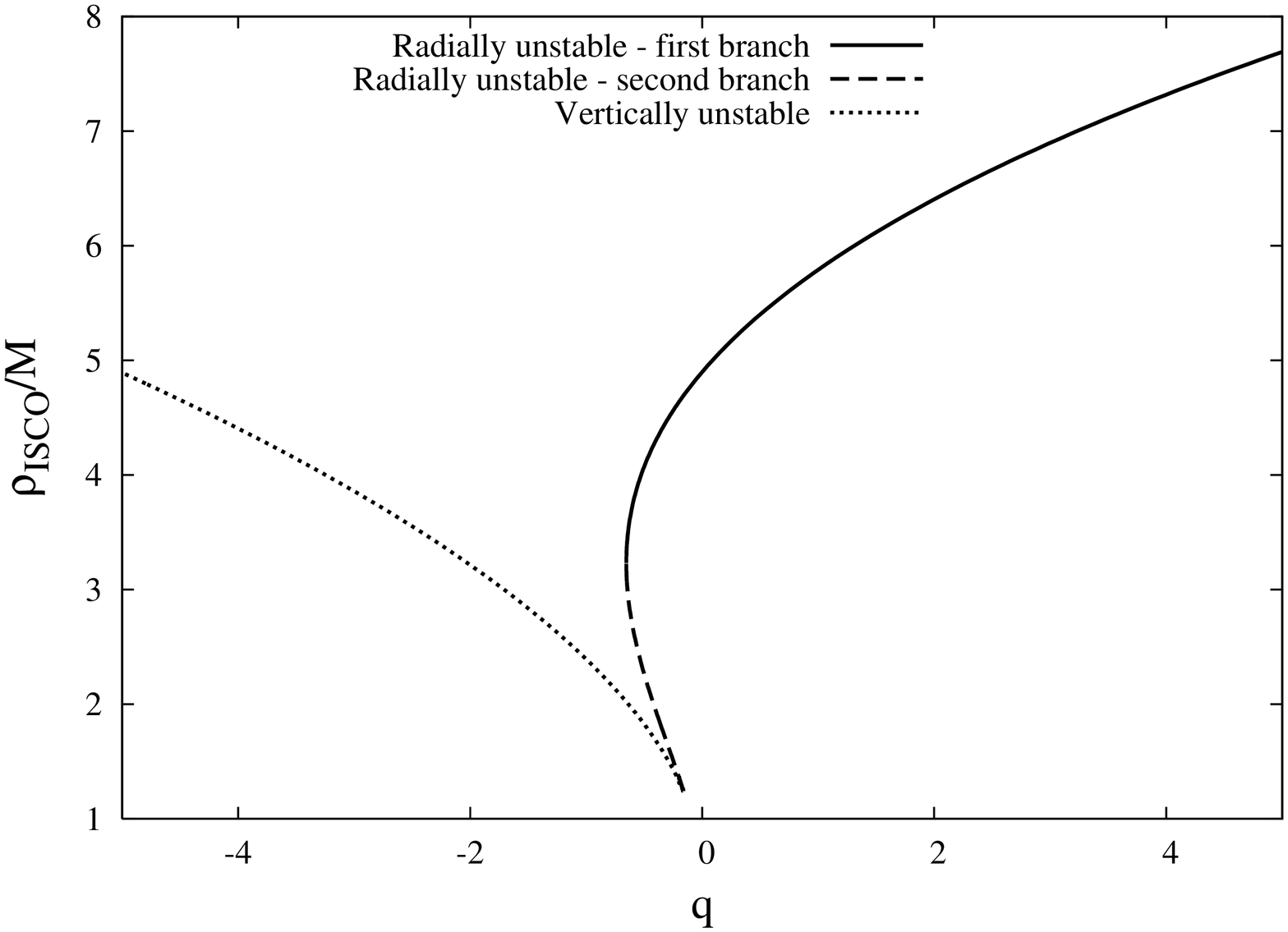} &
\includegraphics[keepaspectratio=true,height=3.5in,angle=-90]{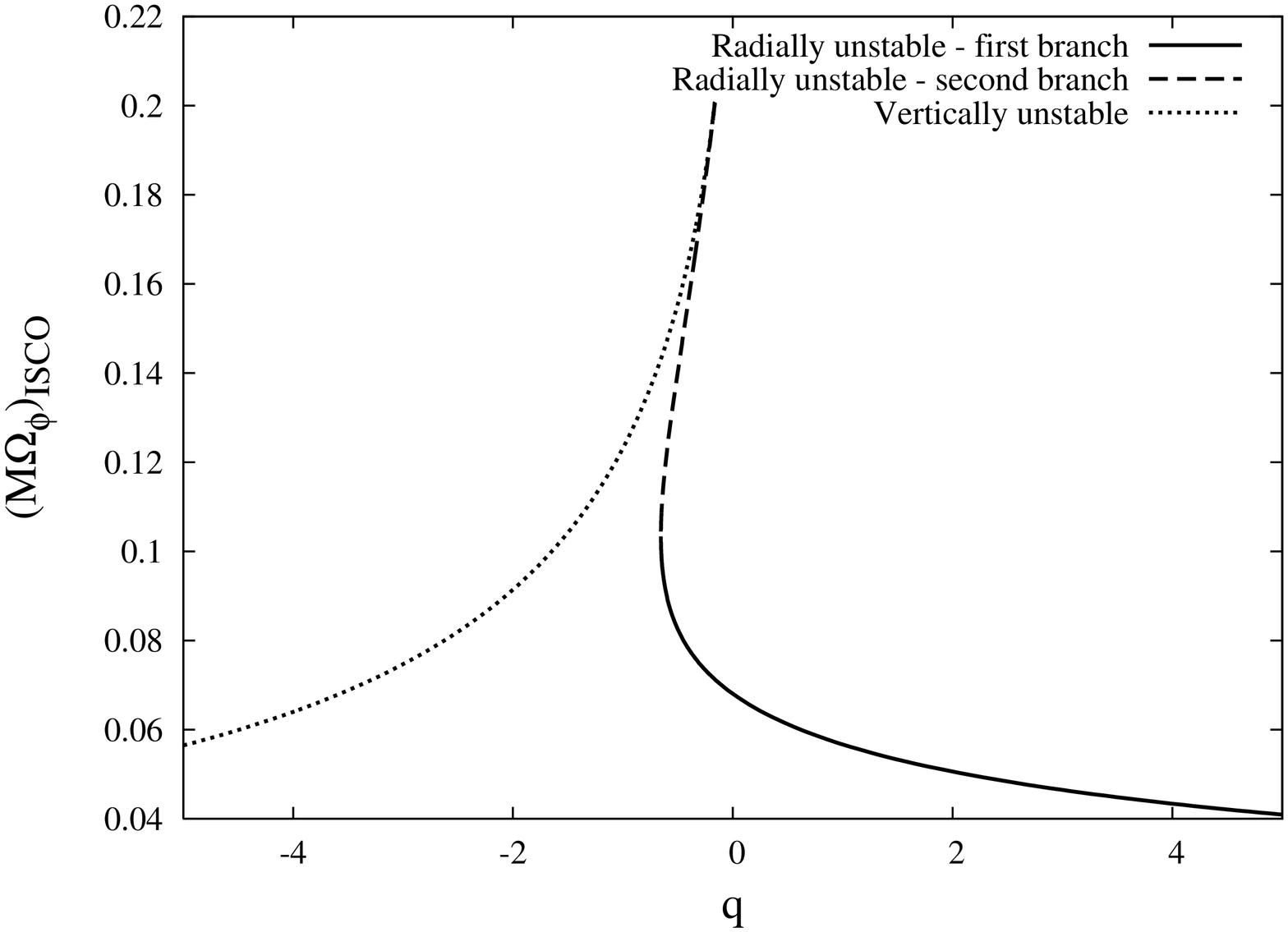} \\
\end{tabular}
\caption{Properties of the equatorial ISCO in spacetimes with $\chi = 0$, as a function of $q$. We show the $\rho$ coordinate of the ISCO (left panel) and the dimensionless frequency of the orbit at the ISCO (right panel). As described in the text, the ISCO radius has three branches, depending on whether it is determined by one of the two branches of radial instability or the branch of vertical instability. These branches are indicated separately in the diagram. For values of $q$ where all three branches are present, the dashed line denotes the ``OSCO'' and the dotted line denotes $\tilde{\rho}_{\rm ISCO}$ as discussed in the text. Allowed orbits lie above the curve in the left panel, and below the curve in the right panel.}
\label{LSOa0}
\end{figure}

\begin{figure}
\centering
\begin{tabular}{cc}
\includegraphics[keepaspectratio=true,height=3.5in,angle=-90]{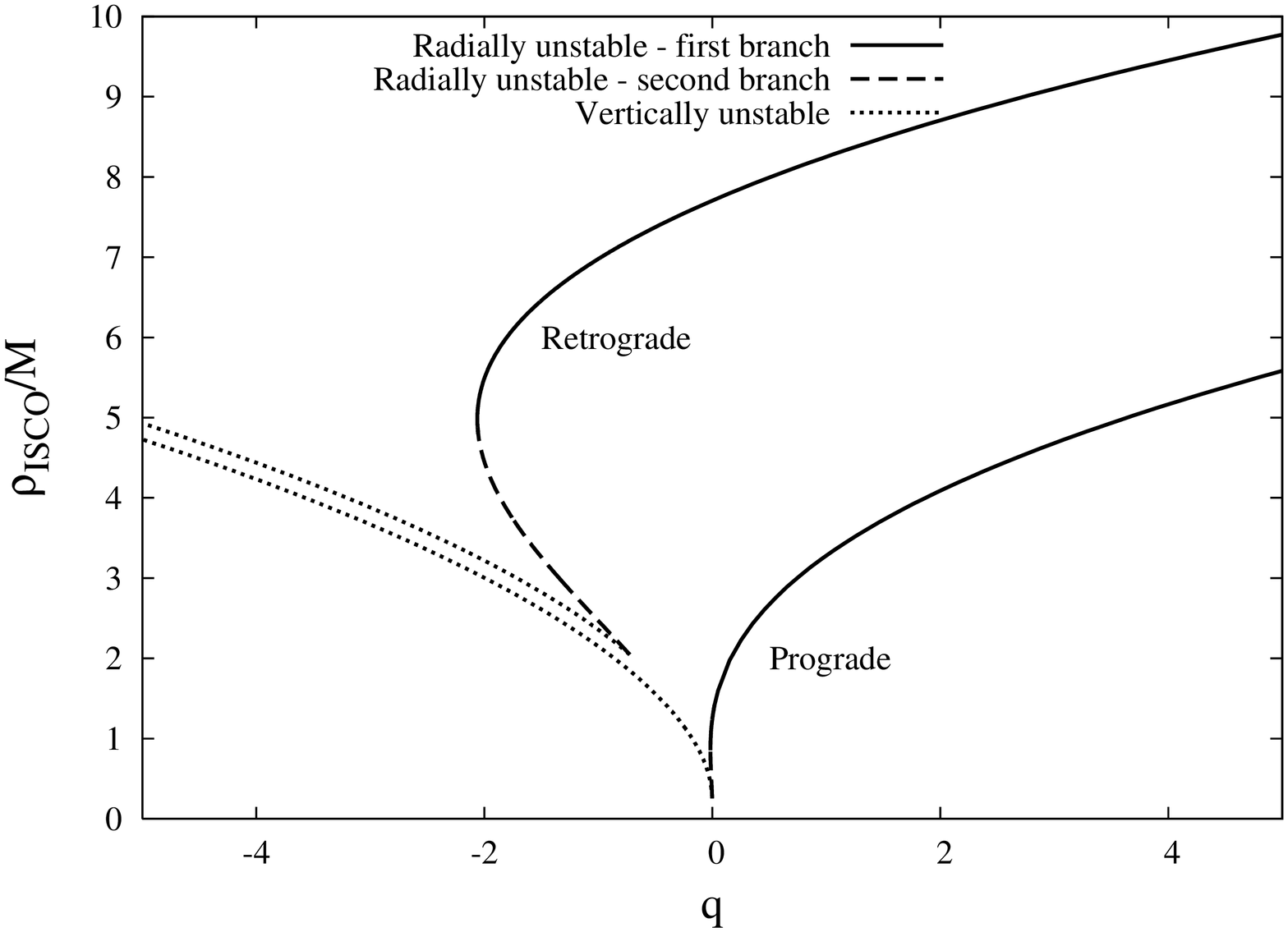} &
\includegraphics[keepaspectratio=true,height=3.5in,angle=-90]{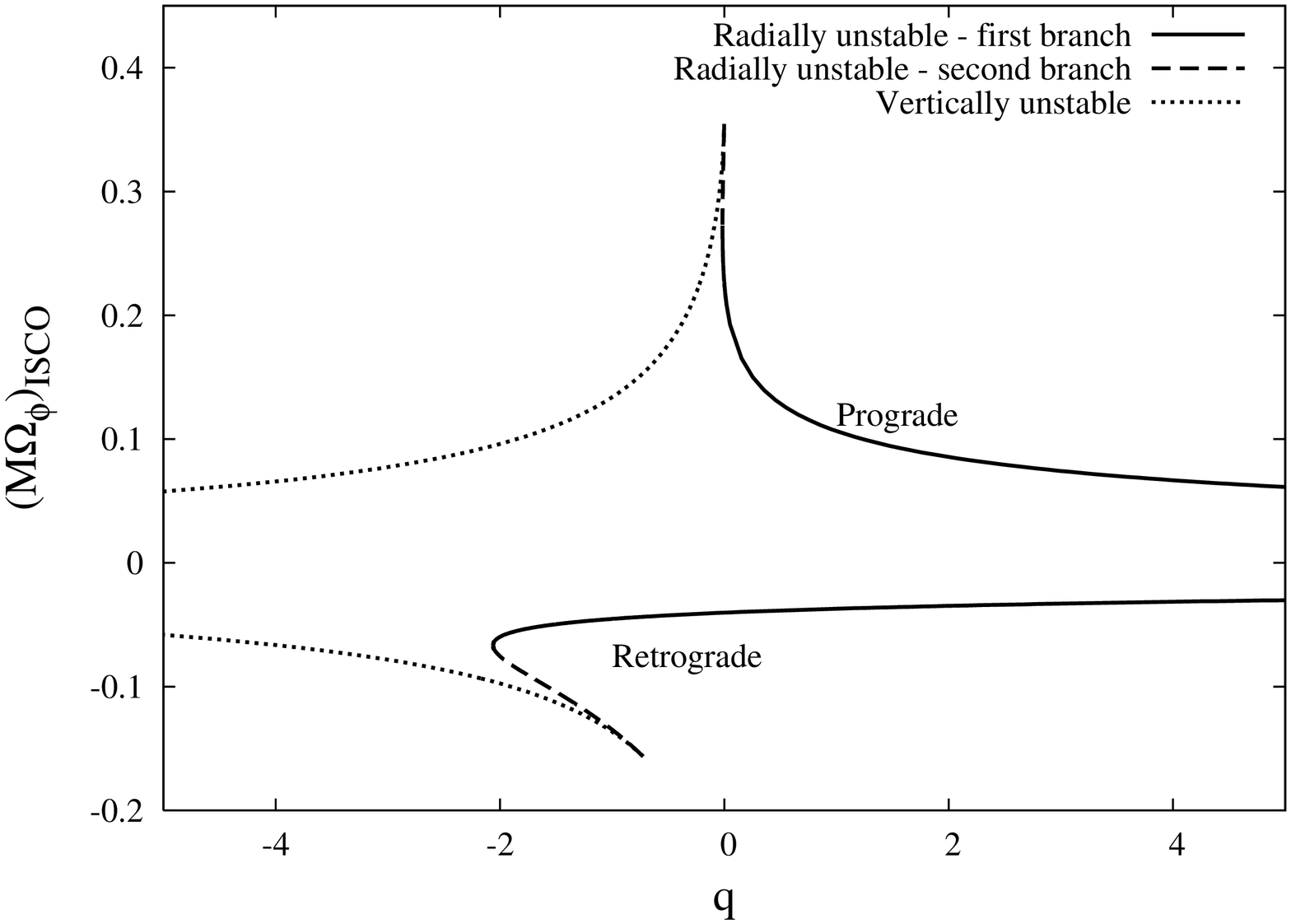} \\
\end{tabular}
\caption{As Figure~\ref{LSOa0}, but now for a spin of $\chi=0.9$. There are now two ISCO curves, one for prograde orbits and one for retrograde orbits. The allowed range of orbital frequencies is given by the region in between the two curves in the right hand panel.}
\label{LSOa09}
\end{figure}

In Figure~\ref{LSOa0} we show the location of the ISCO as a function of $q$ for spacetimes with $\chi=0$. We also show the orbital frequency at the ISCO as a function of $q$, computed using Eq.~\erf{Omegaphi}. For spacetimes with spin, the behavior is qualitatively similar, but there are now two ISCO radii, corresponding to prograde and retrograde orbits respectively. We show results for a spin of $\chi=0.9$ in Figure~\ref{LSOa09}. We note that the ISCO radius is always outside the boundary of the causality-violating region of the spacetime.
For $\chi \neq 0$ and $q > 0$, the ISCO radius is determined by the energy at which the outer allowed region for bound motion (which is a single point for a circular equatorial orbit) merges with the inner allowed region. In that case when the object reached the ISCO it would undergo a transition onto an eccentric/inclined geodesic.

The value of the ISCO frequency depends not only on $q$ but also $M$ and $\chi$. However, as we shall discuss in the next section, it is possible to measure these other parameters using precessions measured when the orbit is in the weak-field. Thus, the ISCO frequency is a powerful probe of the nature of the spacetime since it can be very different even for comparatively small deviations from Kerr. 

\section{Periapsis and orbital-plane precessions}
\label{precession}

In Section~\ref{integrals} we saw that astrophysically relevant orbits in the Manko-Novikov
spacetime are multi-periodic to high precision. In such cases, there is no
smoking-gun signature that indicates the presence of ``bumpiness'' in the
spacetime. Instead, the imprint of the spacetime bumpiness will be
observationally apparent in the location of the last stable orbit, as discussed in the previous section, and in the following ways: (1) in the three fundamental
frequencies of the gravitational waves generated while the inspiraling object is
on an instantaneous geodesic orbit; (2) in the harmonic structure of the
gravitational-wave emission, i.e., the relative amplitudes and phases of the
various harmonics of the fundamental frequencies; and (3) in the evolution of
these frequencies and amplitudes with time as the object inspirals.  A full
analysis of the accuracies that could be achieved in observations would involve
computing gravitational waveforms in the bumpy spacetimes, performing a
Fisher-Matrix analysis to account for parameter correlations, and comparing to
a similar analysis for Kerr. That is beyond the scope of this paper. However,
we can examine the first of these observational consequences by comparing the
fundamental frequencies between the bumpy and Kerr spacetimes.

The complication in such an analysis is to identify orbits between different
spacetimes. Identifying orbits by the $\rho$ and $z$ coordinates is not gauge-invariant, since the meaning of these coordinates depends on the spacetime structure. Identifying orbits via
the energy and angular momentum is gauge-invariant, but these quantities are not directly measurable observationally. However, circular orbits in the
equatorial plane of the spacetime are characterized by a single observable --- the azimuthal frequency of the orbit. We can use this frequency to identify circular equatorial orbits in different spacetimes.

Precession frequencies are absent in exactly
circular equatorial orbits. However, if the circular orbit is perturbed
radially, it will undergo small oscillations at the radial epicyclic frequency,
which is characteristic of the periapsis precession frequency at that radius.
Likewise, if the orbit is perturbed vertically it will undergo small
oscillations at the vertical epicyclic frequency, which is characteristic of
the orbital-plane precession frequency at that radius. We thus 
compare these epicyclic frequencies, as a function of the circular orbital
frequency, between Kerr and bumpy spacetimes.  This comparison was employed
by Ryan, who used it to derive his theorem stating that all
spacetime multipole moments are encoded in the gravitational
waves generated by nearly-circular, nearly-equatorial EMRIs \cite{Ryan1995}. 

An eccentric equatorial orbit can be characterized by two observables --- the orbital frequency and the periapsis precession frequency. These two frequencies can therefore be used to identify orbits in different spacetimes (provided there is an orbit with corresponding frequencies in the Kerr metric). Likewise, the orbital-plane precession frequency can be used to identify inclined orbits between spacetimes\footnote{The `orbital-plane' is not well defined in the strong field. However, we know the gravitational waves should be triperiodic and, in the weak-field, the three periods correspond to the orbital period and the two precessions. When we refer to `orbital-plane precession frequency' we really mean the frequency component of the orbit that corresponds to orbital-plane precession in the weak-field. This will be the frequency of the vertical motion, averaged over many orbits.}. With such an identification, differences in the multipole structure of the spacetime will show up only in the relative amplitudes of the   harmonics and in the evolution of the fundamental frequencies over the inspiral. We will discuss this some more at the end of this section, but a full analysis requires treatment of inspiral in an arbitrary spacetime and is beyond the scope of the current paper.

\subsection{Epicyclic frequencies}

The frequency of epicyclic motion can be derived by perturbing a circular, equatorial orbit in either the
radial or vertical direction. The second order geodesic equations~(\ref{geoalt})
for $z$ and $\rho$ take the form
\begin{eqnarray}
\frac{\rmd}{\rmd \tau} \left(2\,g_{XX}\,\frac{\rmd X}{\rmd \tau}\right) &=& 
\partial_X g_{tt}\,\left(\frac{\rmd t}{\rmd\tau}\right)^2 + 
2 \partial_X g_{t\phi}\,\left(\frac{\rmd t}{\rmd\tau}\right) 
\left(\frac{\rmd \phi}{\rmd\tau}\right) + 
\partial_X g_{\phi\phi}\,\left(\frac{\rmd \phi}{\rmd\tau}\right)^2 \nn \\
&& + \partial_X g_{\rho\rho}\,\left(\frac{\rmd \rho}{\rmd\tau}\right)^2 + 
\partial_X g_{zz}\,\left(\frac{\rmd z}{\rmd\tau}\right)^2.
\end{eqnarray}
Here $X$ denotes either $\rho$ or $z$. The dependence on $\rmd t/\rmd\tau$ and
$\rmd \phi/\rmd\tau$ can be eliminated by using the energy and angular momentum
conservation equations to express these in terms of $E$, $L_z$, $\rho$ and $z$,
as in Eq.~(\ref{tphidot}). Using this form of the equations we can take a
circular, equatorial orbit, $\rho = \rho_c$, $z=0$, and perturb it either in the
radial direction, $\rho = \rho_c + \delta \rho$, $z=0$, or in the vertical
direction, $\rho = \rho_c$, $z=\delta z$. Considering the equations of motion at leading
order in the small orbital perturbation, it is easy to see that the frequencies
of these small epicyclic oscillations are given by
\begin{eqnarray}
\left( \frac{g_{\phi\phi} E-g_{t\phi}L_z}{g_{tt}g_{\phi\phi}-g_{t\phi}^2}
\right)^2\,\Omega_X^2 &=& \nn\\
&&\hspace{-1.5in}\frac{1}{2g_{XX}}\frac{\partial}{\partial X} 
\left( \frac{\partial_X g_{tt} 
\left( g_{\phi\phi}E - g_{t\phi}L_z\right)^2 +2 \partial_X g_{t\phi} 
\left(g_{\phi\phi}E - g_{t\phi}L_z \right)
\left(g_{tt}L_z - g_{t\phi}E \right)} {(g_{tt}g_{\phi\phi}-g_{t\phi}^2)^2}
\right) \nn \\
&&+\frac{1}{2g_{XX}}\frac{\partial}{\partial X} 
\left( \frac{\partial_X g_{\phi\phi} \left(g_{tt}L_z - g_{t\phi}E \right)^2} 
{(g_{tt}g_{\phi\phi}-g_{t\phi}^2)^2}\right)
\end{eqnarray}
As before, $X$ denotes either $\rho$ (for the radial epicyclic frequency
$\Omega_\rho$) or $z$ (for the vertical epicyclic frequency $\Omega_z$). The same result can be derived starting from the effective potential equation
(\ref{Veff}): the frequency $\Omega_X$ is determined by $\partial^2 V_{\rm eff}/\partial X^2$ evaluated at the circular orbit.

%

\subsection{Precessions}

We are interested in precessions rather than the epicyclic
frequency. We define the {\it periapsis precession} as the number of
cycles by which the periapsis advances per radial period (i.e., over one
complete epicyclic oscillation). Likewise, the {\it orbital-plane precession}
is defined as the number of cycles by which the azimuthal
angle to the highest point of the orbit advances during one vertical
oscillation. These precessions, which we denote by $p_X$, 
are related to the epicyclic frequencies, $\Omega_X$, by
\be
p_X = \frac{\Omega_{\phi}}{\Omega_X} - 1.
\ee

The behavior of the precessions can be understood in terms of what happens in the weak-field, far from the black hole, and in the strong-field, close to the ISCO. In the weak-field it is possible to derive expressions for the precessions as functions of the orbital frequency. This was originally done for nearly circular, nearly equatorial orbits by Ryan~\cite{Ryan1995}, who demonstrated that the various spacetime multipole moments enter the precession rate expansion at different orders of $(M\Omega_\phi)^\alpha$. This was the basis for a theorem that, in principle, the weak-field precessions can be used to extract the lowest order spacetime multipole moments. The weak-field expansion of the precessions is summarized in Appendix~\ref{wfprec}. 

In the strong-field, we find that one or the other precession diverges at a certain frequency. This frequency corresponds to the frequency of the ISCO. To understand what is happening, we use the effective potential~\erf{Veff} and consider radial oscillations. For the energy and angular momentum corresponding to the circular equatorial orbit at radius $\rho=\rho_c$, the effective potential in the equatorial plane takes the form $V_{\rm eff} (\rho, z=0) = -\tilde{V}(\rho) (\rho - \rho_-) (\rho - \rho_c)^2$. Here $\tilde{V}(\rho)$ is a function that is strictly positive for $\rho > \rho_-$. The radius $\rho_-$ is the other solution to $V_{\rm eff} (\rho, z=0) = 0$, and $\rho_- < \rho_c$. As the ISCO is approached, the effective potential develops a point of inflection at the location of the turning point rather than a maximum since $\rho_- \rightarrow \rho_c$. The epicyclic frequency for radial oscillations is $\Omega_\rho^2 \propto \tilde{V}(\rho_c) (\rho_c - \rho_-)$, which thus tends to zero as the ISCO is approached. The corresponding periapsis precession diverges. The radius $\rho_-$ corresponds to an unstable circular orbit, and associated with any unstable circular orbit is a bound, eccentric orbit that has an infinite period --- the object comes in from larger radii, and asymptotically approaches the radius of the circular orbit. This is referred to as a ``homoclinic'' orbit, or as a ``zoom-whirl'' orbit in the EMRI literature. As the ISCO is approached, a small perturbation from the location of the circular orbit will put the object onto an orbit that is close to the homoclinic orbit associated with the unstable circular orbit. Hence, it takes a very long time for the object to complete a radial oscillation, but it is moving rapidly in the azimuthal direction the whole time, building up a large periapsis precession.

This understanding leads us to expect the precession to diverge at the location of the ISCO, and this divergence should be like $(\rho_c - \rho_{\rm ISCO})^{-1/2}$, or $(\Omega_{\phi,{\rm ISCO}} - \Omega_\phi)^{-1/2}$. The above argument applies to an ISCO defined by a radial instability (as in the Kerr metric). As we saw in Section~\ref{LSOsec}, the ISCO in the Manko-Novikov spacetimes can be determined by the onset of a vertical instability. In that case, the above argument still applies, but it is now the orbital-plane precession that will diverge as the ISCO is approached. This provides another potential `smoking-gun' for a deviation from the Kerr metric. The divergence in the precession at the ISCO arises as a result of one of the two epicyclic frequencies going to zero. It is these
frequencies that will in principle be observable in the gravitational waves. If an
inspiral is observed starting in the weak-field and up until the last stable orbit (LSO), the
different frequency components could be tracked, and one frequency will
tend to zero as the LSO is approached. This is in principle an observable, and if it is the orbital-plane precession that goes to zero the central body cannot be a Kerr black hole. A more careful treatment of the gravitational-wave emission will be required to understand how practical it will be to make such observations.

\begin{figure}
\includegraphics[keepaspectratio=true,height=5in,angle=-90]{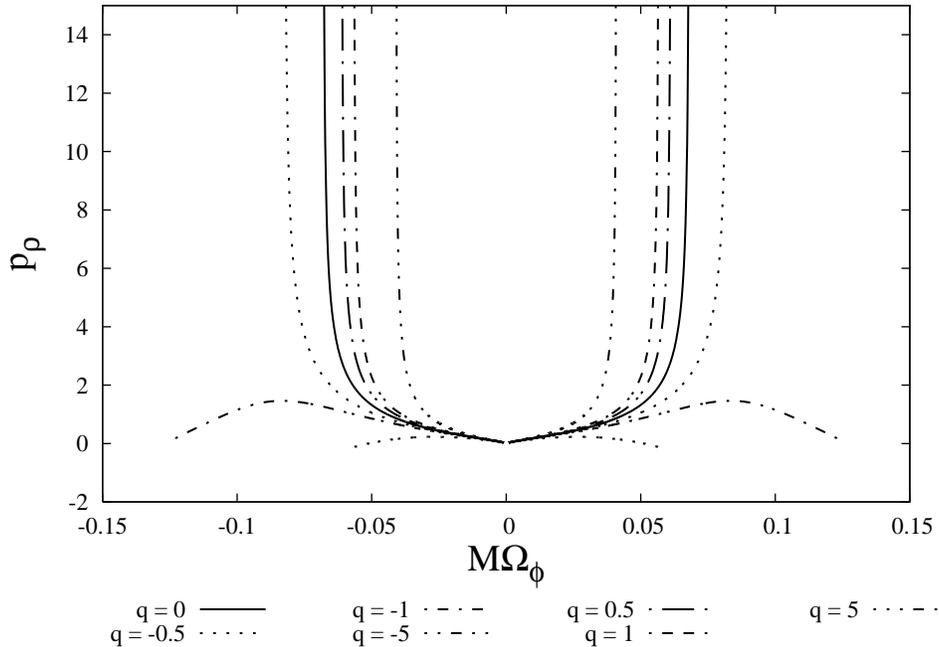}
\caption{Periapsis precession $p_\rho$ versus azimuthal frequency $\Omega_\phi$ 
for $\chi=0$ and various values of $q$.}
\label{fig:perinospin}
\end{figure}

\begin{figure}
\includegraphics[keepaspectratio=true,height=5in,angle=-90]{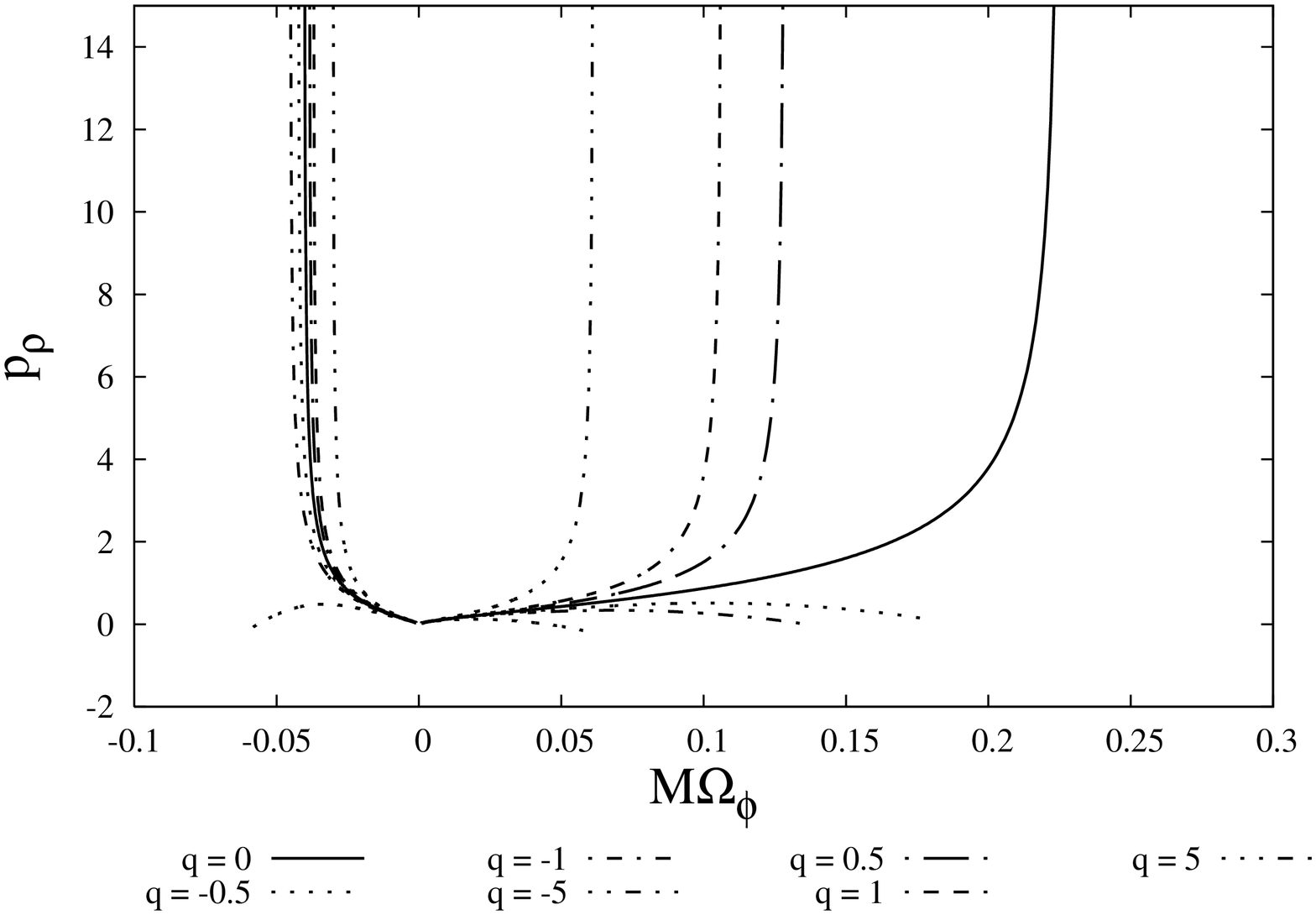}
\caption{Periapsis precession $p_\rho$ versus azimuthal frequency 
$\Omega_\phi$ for $\chi=0.9$ and various values of $q$.}
\label{fig:perispin}
\end{figure}

\begin{figure}
\includegraphics[keepaspectratio=true,height=5in,angle=-90]{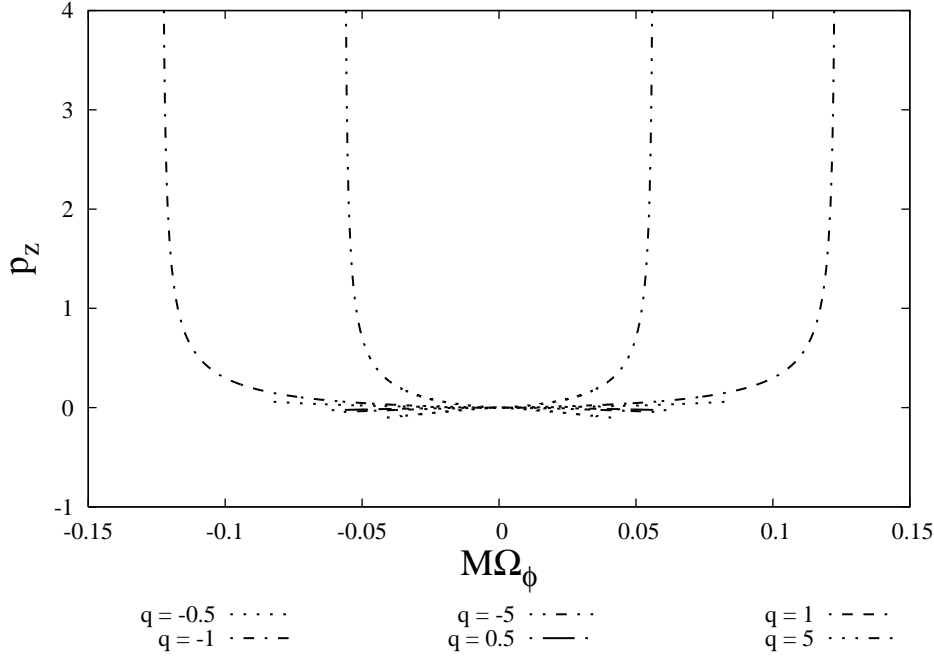}
\caption{Orbital-plane precession $p_z$ versus azimuthal frequency 
$\Omega_\phi$ for $\chi=0$ and various values of $q$.  We do not show 
the case $q=0$ here, since there is no orbital-plane precession in 
Schwarzschild.}
\label{fig:planenospin}
\end{figure}

\begin{figure}
\includegraphics[keepaspectratio=true,height=5in,angle=-90]{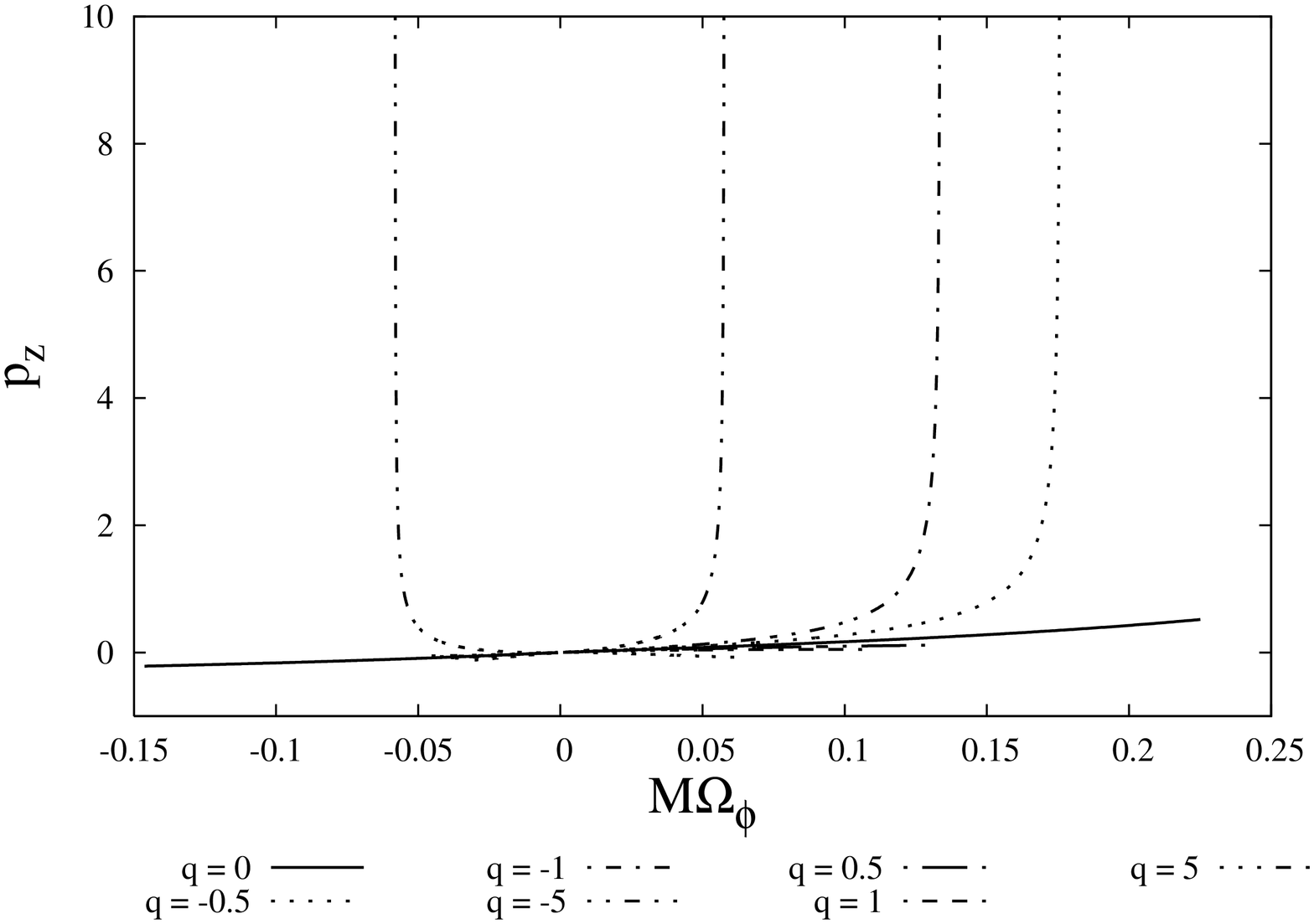}
\caption{Orbital-plane precession $p_z$ versus azimuthal frequency 
$\Omega_\phi$ for $\chi=0.9$ and various values of $q$.}
\label{fig:planespin}
\end{figure}

\begin{figure}
\includegraphics[keepaspectratio=true,height=5in,angle=-90]{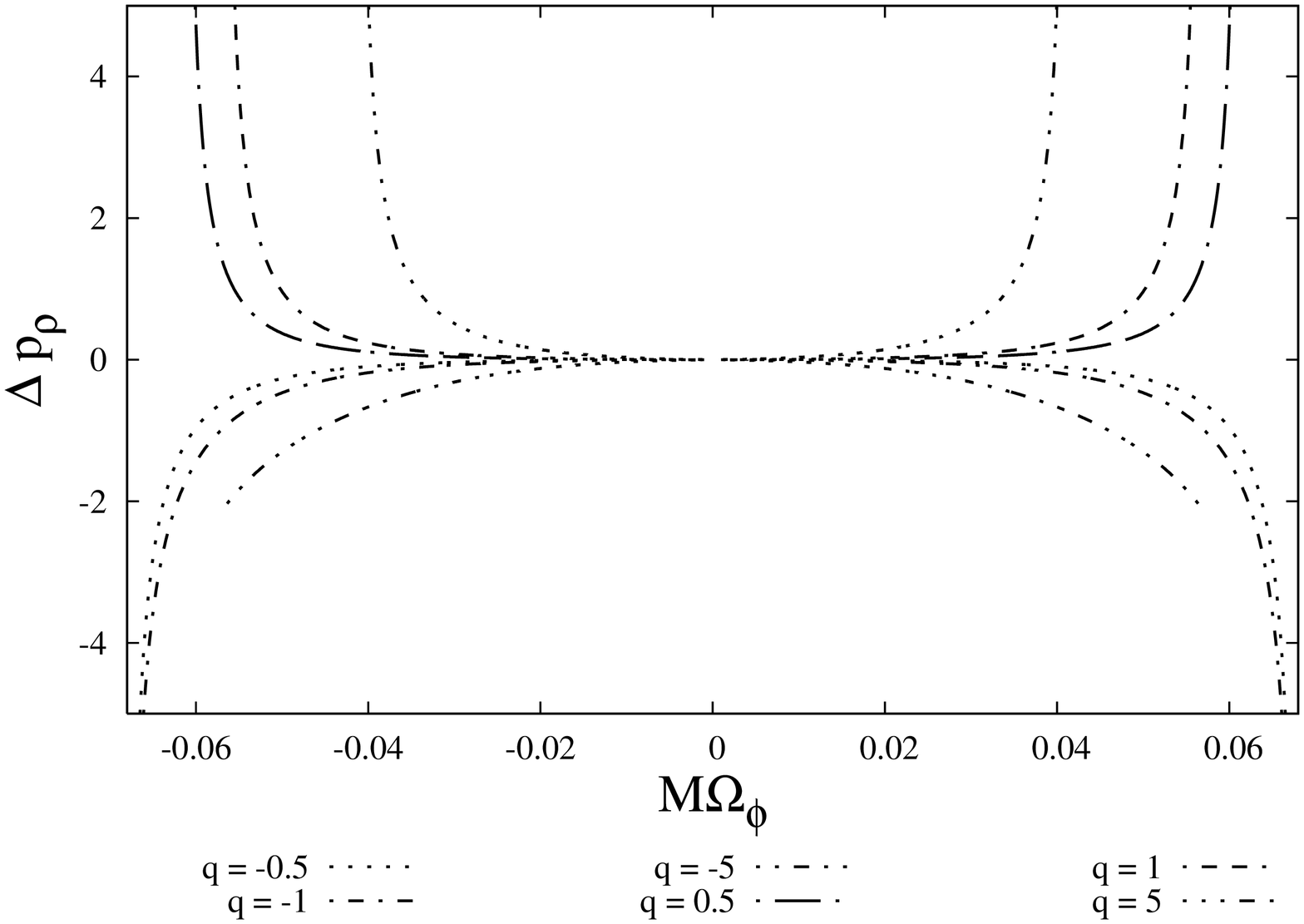}
\caption{Difference between periapsis precessions in a bumpy spacetime with
$\chi=0$ and the Schwarzschild spacetime,
$\Delta p_\rho (\Omega_\phi, q)=p_\rho(\Omega_\phi,q) 
- p_\rho(\Omega_\phi, q=0)$.}
\label{fig:diffperinospin}
\end{figure}

\begin{figure}
\includegraphics[keepaspectratio=true,height=5in,angle=-90]{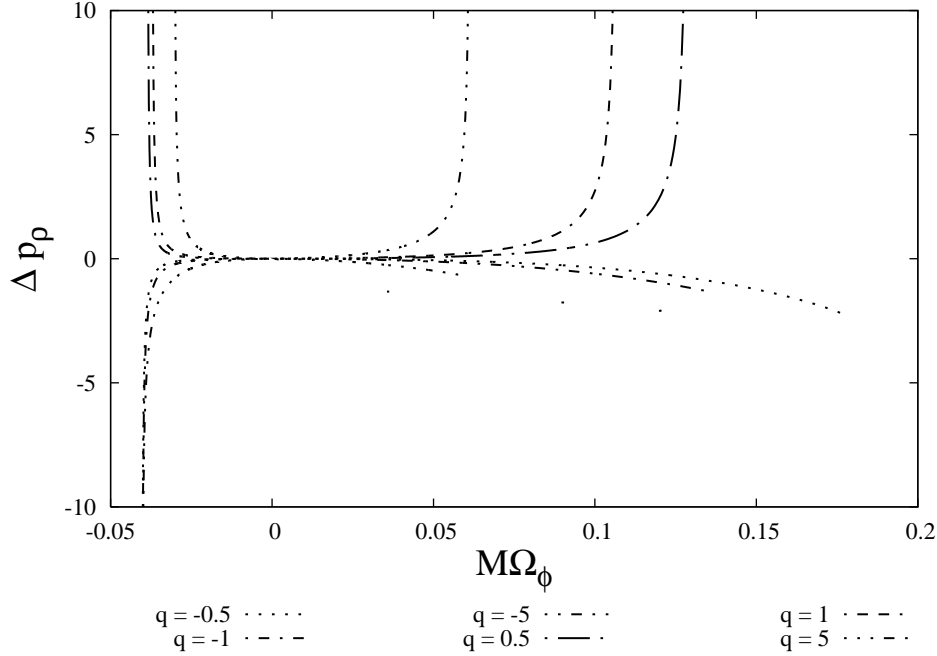}
\caption{Difference between periapsis precessions in a bumpy spacetime with
$\chi=0.9$ and the Kerr spacetime with $\chi=0.9$,
$\Delta p_\rho (\Omega_\phi, q)=p_\rho(\Omega_\phi,q)
- p_\rho(\Omega_\phi, q=0)$.}
\label{fig:diffperispin}
\end{figure}

\begin{figure}
\includegraphics[keepaspectratio=true,height=5in,angle=-90]{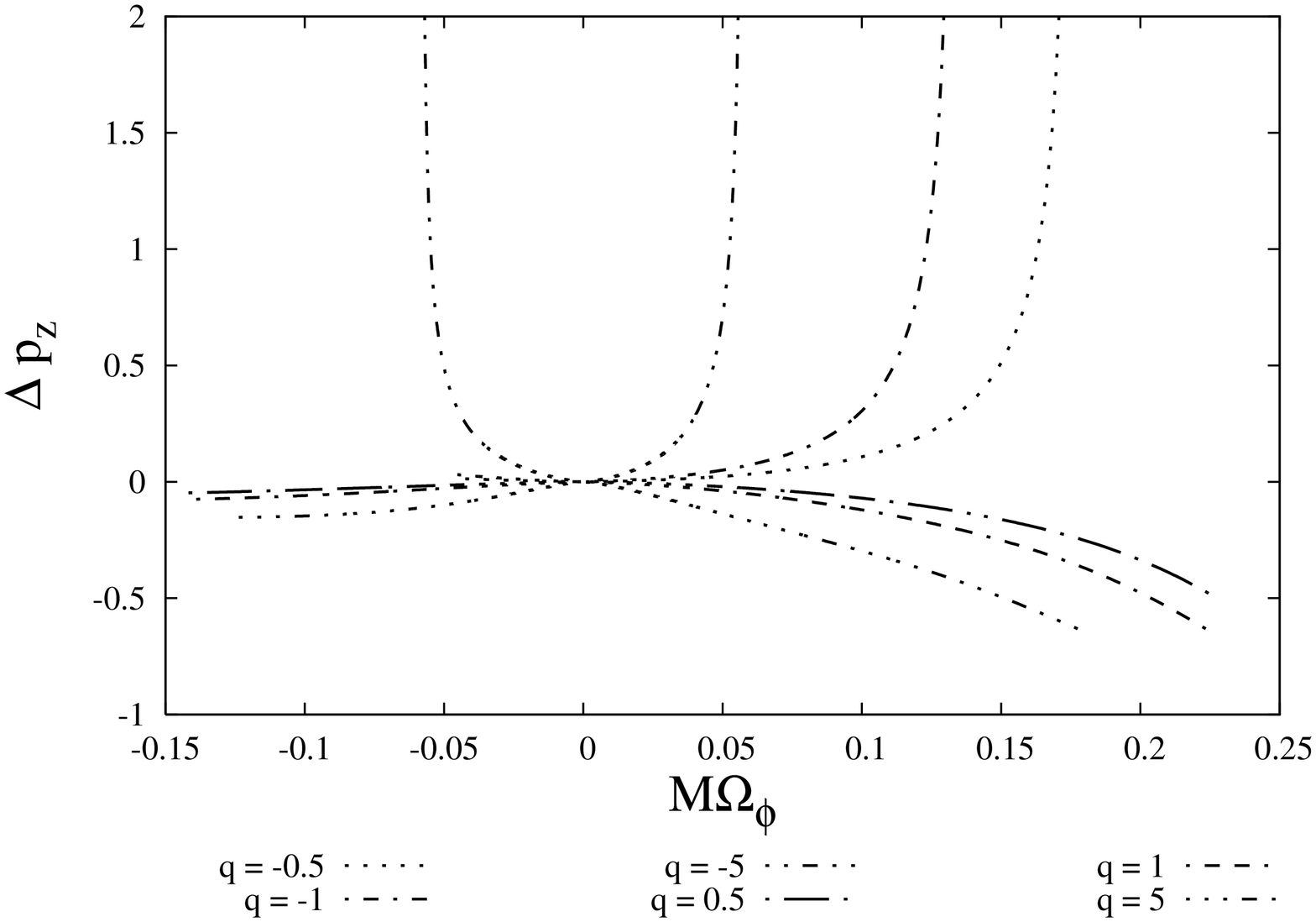}
\caption{Difference between orbital-plane precessions in a bumpy spacetime 
with $\chi=0.9$ and the Kerr spacetime with $\chi=0.9$,
$\Delta p_z (\Omega_\phi, q)=p_z(\Omega_\phi,q)
- p_z(\Omega_\phi, q=0)$.}
\label{fig:diffplanespin}
\end{figure}

In Figures~\ref{fig:perinospin}--\ref{fig:planespin} we show the precessions as a function of $M\Omega_{\phi}$ for a variety of values of $q$. In Figures~\ref{fig:diffperinospin}--\ref{fig:diffplanespin} we present the same results, but now we show the differences between precessions in a bumpy
spacetime with a given $q$ and precessions in the Kerr spacetime with the same
spin parameter $\chi$: $\Delta p_X = p_X - p_X^{\rm Kerr}$.  The variable
$\Delta p_X$  represents the number of cycles of difference, so for instance a
value of $\Delta p_\rho = 0.1$ means that the orbits in the two spacetimes,
although having the same azimuthal frequency, would drift an entire cycle out
of phase in the epicyclic radial oscillation within ten radial orbits. We do not show
results for the difference in the orbital-plane precession for $\chi=0$, since there is no orbital-plane precession in the Schwarzschild spacetime and hence that plot would be identical to Figure~\ref{fig:planenospin}.

Figures~\ref{fig:perinospin} and~\ref{fig:diffperinospin} show the periapsis precession
$r_\rho(\Omega_\phi)$ for $\chi=0$ while Figures~\ref{fig:perispin} and~\ref{fig:diffperispin} show the
periapsis precession for $\chi=0.9$. We see that as the value of $q$ decreases
from zero, the periapsis precession decreases relative to the corresponding
value in the Kerr/Schwarzschild spacetime. By contrast, if $q$ is increased
from zero, the periapsis precession increases. In spacetimes with non-zero
spin, the difference is more extreme for prograde orbits than for retrograde
orbits. This is presumably because retrograde orbits do not get as close to the
central object, and so do not ``feel'' the strong-field deviations in the bumpy
metric. 

For $q\geq -0.5$, the radial epicyclic frequency $\Omega_\rho (\Omega_\phi)$
approaches zero as the ISCO is approached and the periapsis precession $r_\rho$ goes to
infinity for the reasons described above. This is not true of the $q < -0.5$ spacetimes shown, since for those the ISCO is defined by a vertical instability. Figure~\ref{fig:planenospin} shows the orbital-plane precession $r_z(\Omega_\phi)$ for $\chi=0$ and Figures~\ref{fig:planespin} and~\ref{fig:diffplanespin} show the
orbital-plane precession for $\chi=0.9$. As for the case of the periapsis
precession, the orbital-plane precession behaves qualitatively differently
depending on the sign of $q$. The orbital-plane precession is greater for $q<0$
and smaller for $q>0$ compared to the non-bumpy value. As expected, the orbital-plane precession tends to a constant at the ISCO for $q > -0.5$, while it diverges for $q < -0.5$, since the ISCO for the latter spacetimes is defined by a vertical instability as discussed earlier.

Previous authors have looked at precessions in ``bumpy'' spacetimes. As mentioned above, Ryan~\cite{Ryan1995} derived a weak-field expansion for the precessions. Collins \& Hughes~\cite{CH2004} looked at precessions for eccentric equatorial orbits in a perturbed Schwarzschild spacetime, and Glampedakis \& Babak~\cite{Glamp2005} did the same for a perturbed Kerr black hole. However, both pairs of authors did this by comparing orbits with the same coordinates, which is rather unphysical. Our results are consistent with this previous work in the weak-field, but our calculation is the first that can be applied in the strong field, since Ryan's work used a weak-field expansion, and the other work used perturbative spacetimes that break down close to the central body. The behavior in the approach to the ISCO is thus a new result.

It is possible to fit the precessions as a sum of a weak-field expansion (as given in Appendix~\ref{wfprec}) plus a term $A/\sqrt{\Omega_{\phi,{\rm ISCO}} - \Omega_\phi}$. However, only a comparatively few weak-field terms are required to give a good fit, implying that the divergence at the ISCO limits the number of multipole moments that can be recovered from such an expansion. To quantify this statement properly, we must do a full analysis, that includes the effect of inspiral, uses an instrumental noise curve to restrict the observable bandwidth and accounts for parameter correlations via the Fisher Matrix. We can do this by constructing semi-relativistic inspiral waveforms for bumpy spacetimes in the same way that has been used for Kerr inspirals~\cite{GairGlampedakis2006,Babak2007}. This is beyond the scope of the present paper. However, there are several things that we can take away from the current results --- the location of the ISCO has a strong influence on precessions that could be observable, in particular the nature of the instability that defines the ISCO could be a clear indicator of a non-Kerr system; precessions can be very different in the strong field in the presence of a deviation; circular orbits with frequencies very different from the Kerr value exist in some bumpy spacetimes, so another observable signature would be that an inspiral persists at frequencies inside the Kerr ISCO.

\subsection{Effect of eccentricity}
As discussed above, the measurement of the precessions as a function of orbital frequency for nearly circular, nearly equatorial orbits would in principle allow measurement of the spacetime multipole moments~\cite{Ryan1995}. In practice, however, the precessions will only be manifest in the observed gravitational waves if the orbit is not circular and equatorial, so we need to understand how the dependence of the precessions on azimuthal frequency differs when we relax the assumption of near-circularity. In the following, we shall focus on the periapsis precession of eccentric but equatorial orbits.

The eccentricity of an orbit modifies two things --- 1) the frequency associated with the periapsis precession as a function of the orbital frequency; 2) the relative amplitudes of different harmonics of these two frequencies in the observed GWs. To accurately compute the dependence of the harmonic structure on eccentricity for a generic orbit, we need to know details of GW generation in a spacetime with arbitrary multipole moments. This is a difficult problem, so we focus on the effect of eccentricity on the periapsis precession frequency itself.
We consider an eccentric equatorial orbit in the Kerr spacetime, and use $\Omega_\phi$ to denote the average azimuthal frequency (i.e., the total advance in $\phi$ over one radial period, divided by the period of the radial motion). We define an orbital eccentricity, $e$, such that the ratio of the Boyer-Lindquist radii of the periapsis, $r_p$, and apapsis, $r_a$, of the radial motion is $r_p/r_a = (1-e)/(1+e)$. With these definitions, the periapsis precession as defined above can be found to be
\ba
p_{\rho} &=& 3 \left(\frac{M\Omega_\phi}{(1-e^2)^{\frac{3}{2}}}\right)^{\frac{2}{3}} -4\chi \left(\frac{M\Omega_\phi}{(1-e^2)^{\frac{3}{2}}}\right) +\frac{3(18-7e^2+2\chi^2)}{4} \left(\frac{M\Omega_\phi}{(1-e^2)^{\frac{3}{2}}}\right)^{\frac{4}{3}}  \nonumber \\ && -(34-18e^2)\,\chi\, \left(\frac{M\Omega_\phi}{(1-e^2)^{\frac{3}{2}}}\right)^{\frac{5}{3}} + \cdots \label{Kerreccprec}
\ea
where we are expanding in the weak field, $M\Omega_{\phi} \ll 1$. The corresponding result for a spacetime with an excess quadrupole moment can be found at lowest order by replacing the term in $\chi^2$ with $\chi^2+q$, since the quadrupole moment of a Kerr black hole is $\chi^2$ as discussed earlier. 

In the circular limit, $e=0$, the expansion~\erf{Kerreccprec} allows us to extract $M$ from the coefficient of $\Omega_{\phi}^{2/3}$, $\chi$ from the coefficient of $\Omega_\phi$, $q$ from the coefficient of $\Omega_\phi^{4/3}$ etc. However, if we expand to lowest order in the eccentricity, $e$, it is clear that the effect of a small excess oblate quadrupole moment $q > 0$ could be mimicked, at leading order, by an eccentricity evolving as $e^2 = 2(M\Omega_\phi)^{-2/3} q$. The two possibilities are then distinguished by knowing how the eccentricity should evolve with $M\Omega_{\phi}$.

The expansion~\erf{Kerreccprec} contains redundancy, since the coefficient of $M\Omega_{\phi}^{5/3}$ also depends only on the lowest current moment, $\chi$. If the eccentricity of the orbit did not evolve with time the first four terms in the expansion would determine $M$, $\chi$, $q$ and the eccentricity $e$, and higher terms would determine the remaining multipole moments as in the circular case. However, the eccentricity does evolve with time. In practice, we will only observe EMRIs as they evolve through a finite range of frequencies (determined by the detector sensitivity). During that period, the evolution will be driven entirely by gravitational-wave emission. This means that we can quantify the eccentricity of the orbit by a single number --- the periapsis at which the eccentricity was equal to $1$ if we integrated the inspiral backwards in time, assuming a purely GW driven inspiral. Specifying this parameter {\it and the multipole structure of the spacetime} determines the eccentricity as a function of $M\Omega_\phi$. Determining this relationship, however, requires knowing the details of GW emission in an arbitrary spacetime. 

A complication arises because the ratio $M\Omega_\phi/(1-e^2)^{3/2}$ tends to a constant at the point where $e=1$. Assuming that this occurred in the weak-field, $M\Omega_\phi/(1-e^2)^{3/2} \ll 1$, this can be seen by considering the leading order term in $\rmd e/\rmd \Omega_\phi$ in the weak-field (see for instance~\cite{GairGlampedakis2006})
\be
\frac{\rmd e}{\rmd (M\Omega_\phi)} = \frac{-(304+121e^2) (1-e^2) e}{3(M\Omega_\phi)(96+292e^2+37e^4)} .
\ee
Denoting $X=1-e^2$ and expanding in the limit $M\Omega_\phi \rightarrow 0$, $X \rightarrow 1$, we find
\be
\frac{\rmd X}{\rmd (M\Omega_\phi)} \approx \frac{2}{3} \frac{X}{M\Omega_\phi} \Rightarrow X = X_0 (M\Omega_{\phi})^{\frac{2}{3}} + O(\Omega_{\phi}^{\frac{4}{3}}) \label{asymbehav}
\ee
in which $X_0$ is a constant that is related to the periapse at ``capture'' when $X=0$. If the capture occurs in the strong field, the ratio $M\Omega_\phi/(1-e^2)^{3/2}$ would still tend to a constant if we integrated backward until $e\to 1$. Although the inspiral would not be observed as $e \to 1$, that section of the inspiral does affect the portion that we can observe.

We now substitute the asymptotic behavior~\erf{asymbehav} into Eq.~\erf{Kerreccprec}, to obtain an expansion of the periapsis precession as a function of the angular frequency in the form $p_{\rho} = a_0 + a_2 (M\Omega_\phi)^{2/3} + a_3 (M\Omega_\phi) + \cdots$, where $a_0, a_2, a_3$ etc. are constants. In contrast to the circular case, each of these coefficients depends on all the spacetime multipole moments, so multipole extraction from the periapsis precession expansion is no longer straightforward. The reason for this qualitative difference between circular and eccentric orbits is that it is only possible to observe an eccentric inspiral over a finite range of periapsis, since the orbit is captured with a certain finite periapsis, while a circular orbit could inspiral from infinity. The various multipole moments have different radial dependencies, thus if one can observe the precession frequency at any radius it makes sense that all the moments can be separately extracted, while this is more difficult if only a finite section of the spacetime is explored. 

In practice, this difficulty also arises when observing a circular inspiral, since the radiation can only be detected in a certain frequency range. One can parameterize an observation by the frequency at the start of the observation, $\Omega_0 = \Omega_\phi(t=0)$. A Taylor series expansion of the precession (see Eq.~\erf{MNweakprec} in the Appendix) then gives
\ba
p_{\rho} &=& \left(3 \left(M\Omega_0\right)^{\frac{2}{3}} -4\chi \left(M\Omega_0\right)+\frac{3}{2}\left(9+\chi^2 + q\right)  \left(M\Omega_0\right)^{\frac{4}{3}} + \cdots \right) \nonumber \\
&& \hspace{0.1in} +  \left(2 \left(M\Omega_0\right)^{\frac{2}{3}} -4\chi \left(M\Omega_0\right)+2\left(9+\chi^2 + q\right)  \left(M\Omega_0\right)^{\frac{4}{3}} + \cdots \right) \frac{\Omega_\phi - \Omega_0}{\Omega_0} \nonumber \\
&& \hspace{0.1in} +  \left(-\frac{1}{3} \left(M\Omega_0\right)^{\frac{2}{3}} +\frac{1}{3}\left(9+\chi^2 + q\right)  \left(M\Omega_0\right)^{\frac{4}{3}} + \cdots \right) \left(\frac{\Omega_\phi - \Omega_0}{\Omega_0}\right)^2 + \cdots \nonumber \\
&=& b_0 + b_1  \frac{\Omega_\phi - \Omega_0}{\Omega_0} + b_2 \left(\frac{\Omega_\phi - \Omega_0}{\Omega_0}\right)^2 + \cdots
\ea
In this kind of expansion the multipole moments again contribute at all orders. However, provided the initial frequency $M\Omega_0 \ll 1$, the dominant piece of the constant term, $b_0$, is $\left(M\Omega_0\right)^{\frac{2}{3}}$, so this term can be used to estimate $M$. Similarly, the dominant piece of $2b_0 - 3b_1$ is $4\chi \left(M\Omega_0\right)$, so this can be used to estimate $\chi$, and that estimate of $\chi$ can be used to improve the estimate of $M$ from $b_0$. The dominant piece of $b_0 -b_1 +3b_2$ is $
\left(9+\chi^2 + q\right)/2  \left(M\Omega_0\right)^{\frac{4}{3}}$, so this can be used to estimate the excess quadrupole moment $q$ and so on. In the same way, if an eccentric inspiral is observed in a regime where the initial frequency is small (and hence the frequency at capture was also small), we can use the same type of expansion and use combinations of the coefficients to successively extract each multipole moment and the initial eccentricity. To do this requires an expansion of $e^2 - e_0^2$ as a function of $\Omega_\phi/\Omega_0 - 1$. The necessary derivatives $\rmd e^2/\rmd (M\Omega_\phi)$ are known in the weak-field, and to lowest order in the multipoles (see, for example, reference~\cite{GairGlampedakis2006}). However, this calculation is somewhat involved and beyond the scope of this paper.

The above discussion indicates that the periapsis precession rate can be used on its own to measure the multipole moments from an eccentric equatorial inspiral, although this is more difficult than for the circular equatorial case. However, the value of the precession is not the only observable. As mentioned earlier, the relative amplitude of the various harmonics of the orbital frequencies is a powerful probe of the orbital eccentricity. To exploit this harmonic structure we also need to know how the amplitudes of the harmonics depend on the spacetime multipole moments. However, if the deviations from the Kerr metric multipole structure are small, we could imagine using the Kerr harmonic amplitude relation to estimate the eccentricity (and inclination) of the orbit, and then use the precessions to extract the multipole structure. Proper insight can be gained using the approximate semi-relativistic waveforms described earlier or post-Newtonian expansions of the gravitational waveforms. Such an investigation will be an important extension of the current work.


\section{Summary}
\label{future}
In this paper we have discussed various observational signatures that could leave an imprint on an EMRI gravitational waveform if the spacetime in which the EMRI was occurring deviated from the Kerr metric. We have seen that some orbits in ``bumpy'' spacetimes lack a fourth integral of the motion, and appear ergodic. Geodesics in the Kerr spacetime have a complete set of integrals, so if an apparently ergodic orbit was observed it would be a clear signature of a non-Kerr central object. However, regions of ergodic motion only appear very close to the central object, in a regime which is probably inaccessible to a star inspiraling from large distances. Most astrophysically relevant orbits are regular and appear to possess an approximate fourth integral of the motion, and the orbits are tri-periodic to high accuracy. The deviations of the central body from Kerr then manifest themselves only in the changes in the three fundamental frequencies of the motion and the relative amplitude of the different harmonics of these frequencies present in the gravitational waves. For nearly circular, nearly equatorial orbits, the dependence of the precession frequencies on the orbital frequency is well fit by a combination of a weak field expansion that encodes the multipole moments at different orders, plus a term that diverges as the innermost stable circular orbit is approached. The frequency of the ISCO and its nature (whether it is defined by a radial or vertical instability) is another observable signature of a non-Kerr central object.

To derive these results, we have focussed on a particular family of spacetimes due to Manko and Novikov~\cite{MN1992}. However, we expect the generic features of the results in the weak field and as the ISCO is approached to be true for a wide range of spacetimes. Chaos has been found for geodesic motion in several different metrics by various authors~\cite{ssm96,LV97,GL01,Gueron2002,Dubeibe2007}. In all cases, however, the onset of chaos was qualitatively similar to what we found here --- it occurred only very close to the central object, and for a very limited range of orbital parameters. The conclusion that gravitational waves from ergodic EMRIs are unlikely to be observed is thus probably quite robust.

Precessions for spacetimes that deviate from the Kerr metric have also been considered by several authors~\cite{Ryan1995,CH2004,Glamp2005}. Our results agree with this previous work in the weak-field as they should. However, the results in the present paper are the first that are valid in the strong-field since previous work was either based on a weak-field expansion~\cite{Ryan1995} or a perturbative spacetime~\cite{CH2004,Glamp2005}. The main feature of the precessions in the strong-field --- the divergence of one of the precessions as the ISCO is approached --- is expected from spacetime-independent considerations and therefore should be a general feature of inspirals in any spacetime. The present work, and earlier research~\cite{CH2004,Glamp2005}, has also considered only solutions that first differ from the Kerr metric in the mass quadrupole moment. The Manko-Novikov solutions~\cite{MN1992} include spacetimes that first differ at higher orders. While we have not considered such solutions, we expect the generic features to be similar. The precessions will be closer to the Kerr values for a greater fraction of the inspiral, and the ISCO will be at a different frequency, but the qualitative behavior in the approach to ISCO should be the same.

The next step in understanding how gravitational-wave detectors might identify non-Kerr central objects from EMRI observations is to consider the gravitational waveforms produced during an inspiral. Any analysis should account for both parameter correlations and the finite bandwidth and observation time of gravitational-wave detectors by using a Fisher Matrix analysis. Glampedakis and Babak~\cite{Glamp2005} constructed approximate gravitational waveforms generated by orbits in a perturbed Kerr spacetime, but they considered only waveforms from geodesics (i.e., not inspirals) and compared waveforms with the same orbital parameters. These are not observable quantities (unlike the frequency of the orbit which we used as a basis for comparison here) and such a calculation does not account for parameter correlations. Barack and Cutler~\cite{BC2007} did a full Fisher Matrix analysis of this problem, and estimated that a LISA observation of an EMRI could measure the quadrupole moment of a body to an accuracy of $10^{-3}$ while simultaneously measuring the mass and spin to $10^{-4}$. That calculation was based on an approximate waveform model devised to describe Kerr inspirals. The expressions governing the inspiral were modified by adding the leading order effect of a quadrupole moment to the energy and angular momentum fluxes. The waveform generation part of the algorithm was left unchanged. Although this result is a good guide, the calculation contained a number of inconsistencies. For Kerr inspirals, semi-relativistic ``kludge'' waveforms based on combining exact geodesic motion with approximate gravitational-wave emission formulae have proven to give accurate results~\cite{GairGlampedakis2006,Babak2007}. The same method could be used to produce waveforms for inspiral in the Manko-Novikov spacetimes, by changing the geodesic equations and augmenting the inspiral fluxes appropriately. Such an approach will not generate totally accurate gravitational waveforms, but it will reproduce the main features of the orbit --- the precession frequencies, the orbital shape and the frequency of the ISCO. A study of gravitational waveforms generated in ``bumpy'' spacetimes will provide useful guidance for future detectors such as to what precision an observation could determine that an inspiral is an inspiral into a Kerr black hole and how well observations can distinguish different types of deviation from Kerr, e.g., an exotic central object from a naked singularity from a Kerr black hole with external matter.
 

\begin{acknowledgments} We thank Kip Thorne, Jeandrew Brink, Geoffrey
Lovelace and Hua Fang for useful discussions and Scott Hughes for useful comments on the manuscript.  JG's work was supported by St.~Catharine's College, Cambridge.  IM thanks the Brinson Foundation, NASA grant NNG04GK98G and NSF grant PHY-0601459 for financial support. CL thanks the Moore Foundation for support.
\end{acknowledgments}

\appendix
\section{Chaotic Motion in Newtonian Gravity}
\label{newtchaos}
\begin{figure}
\centering
\begin{tabular}{cc}
\includegraphics[keepaspectratio=true,height=3.5in,angle=-90]{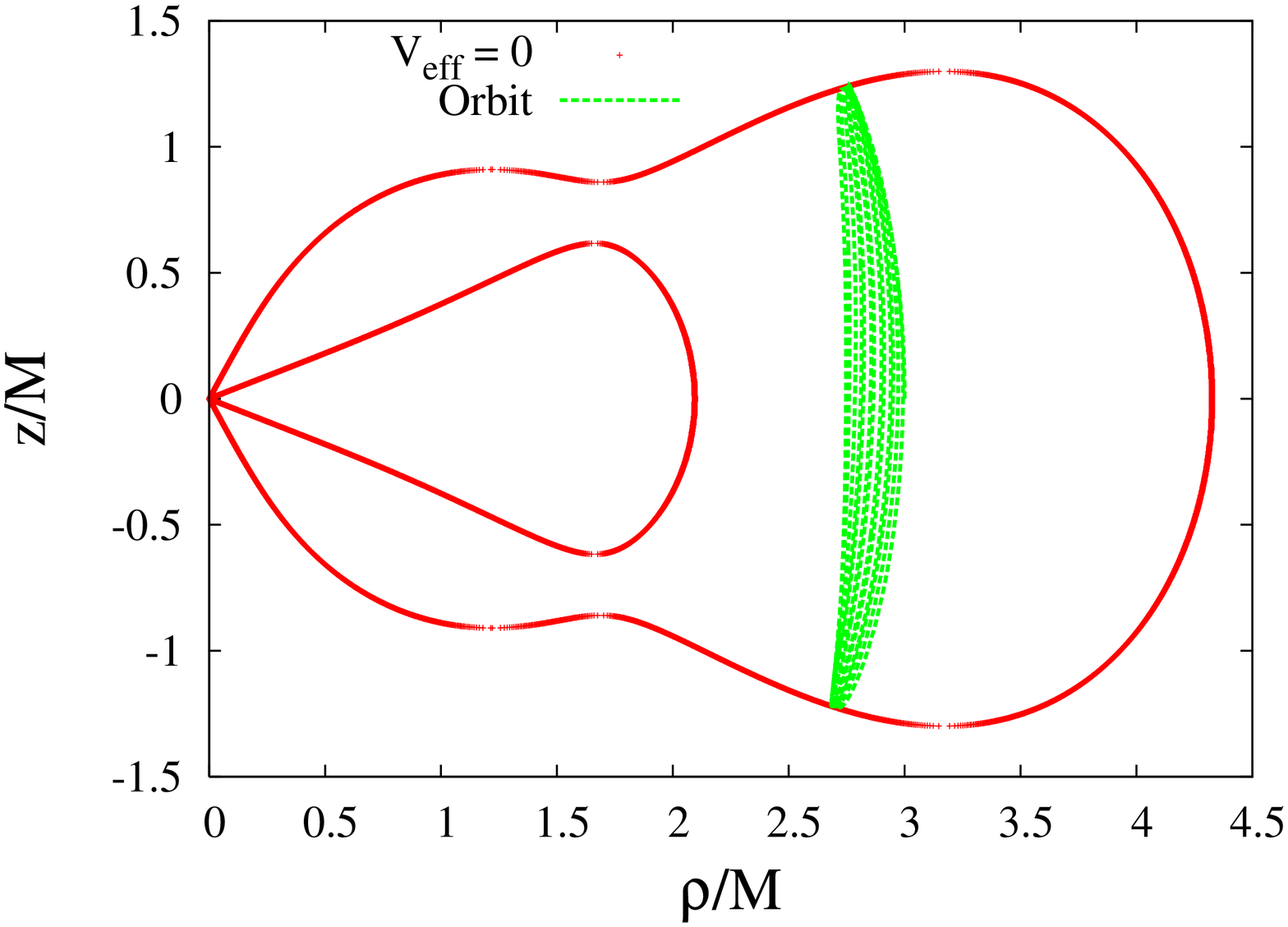} &
\includegraphics[keepaspectratio=true,height=3.5in,angle=-90]{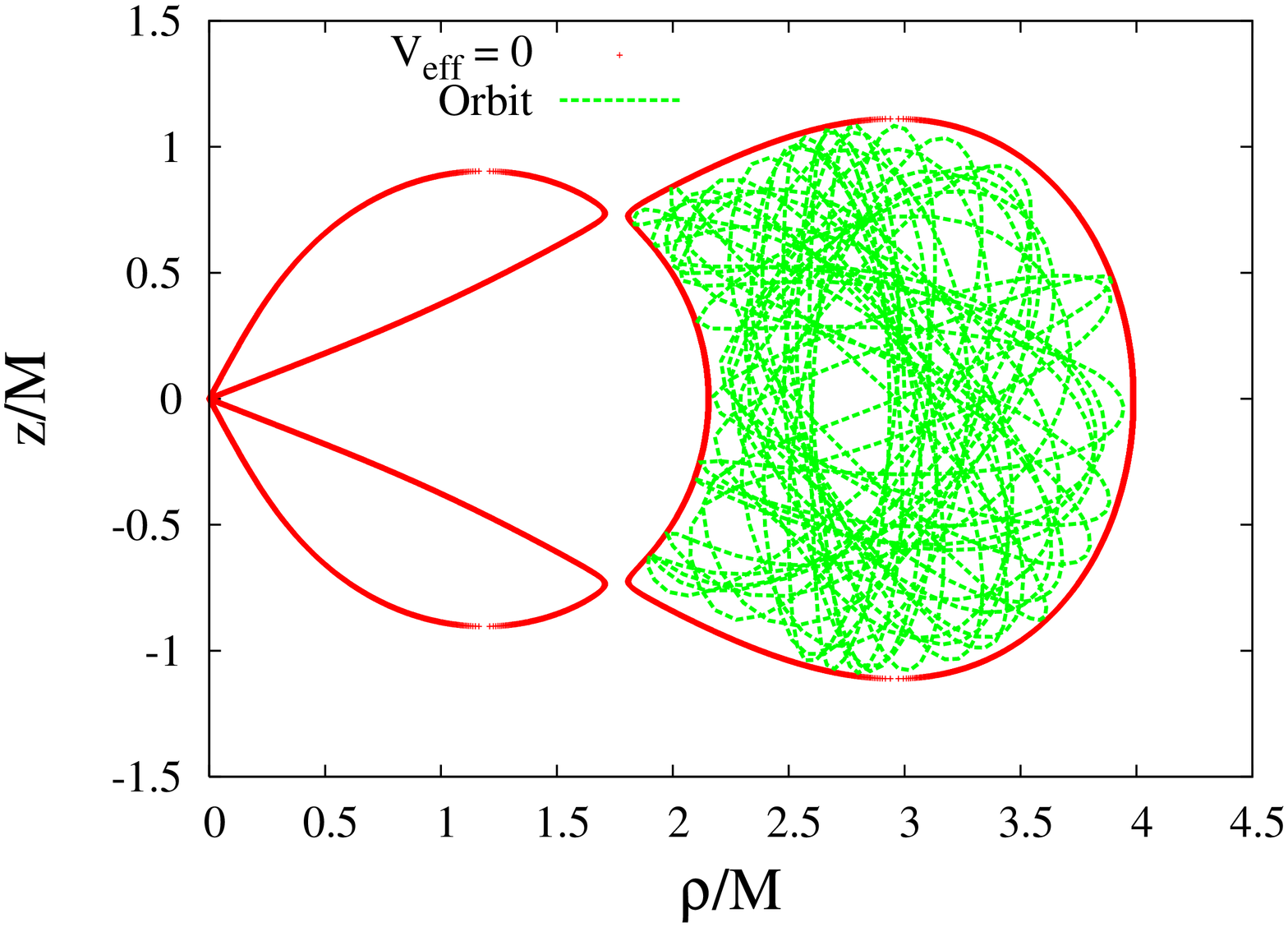} \\
\includegraphics[keepaspectratio=true,height=3.5in,angle=-90]{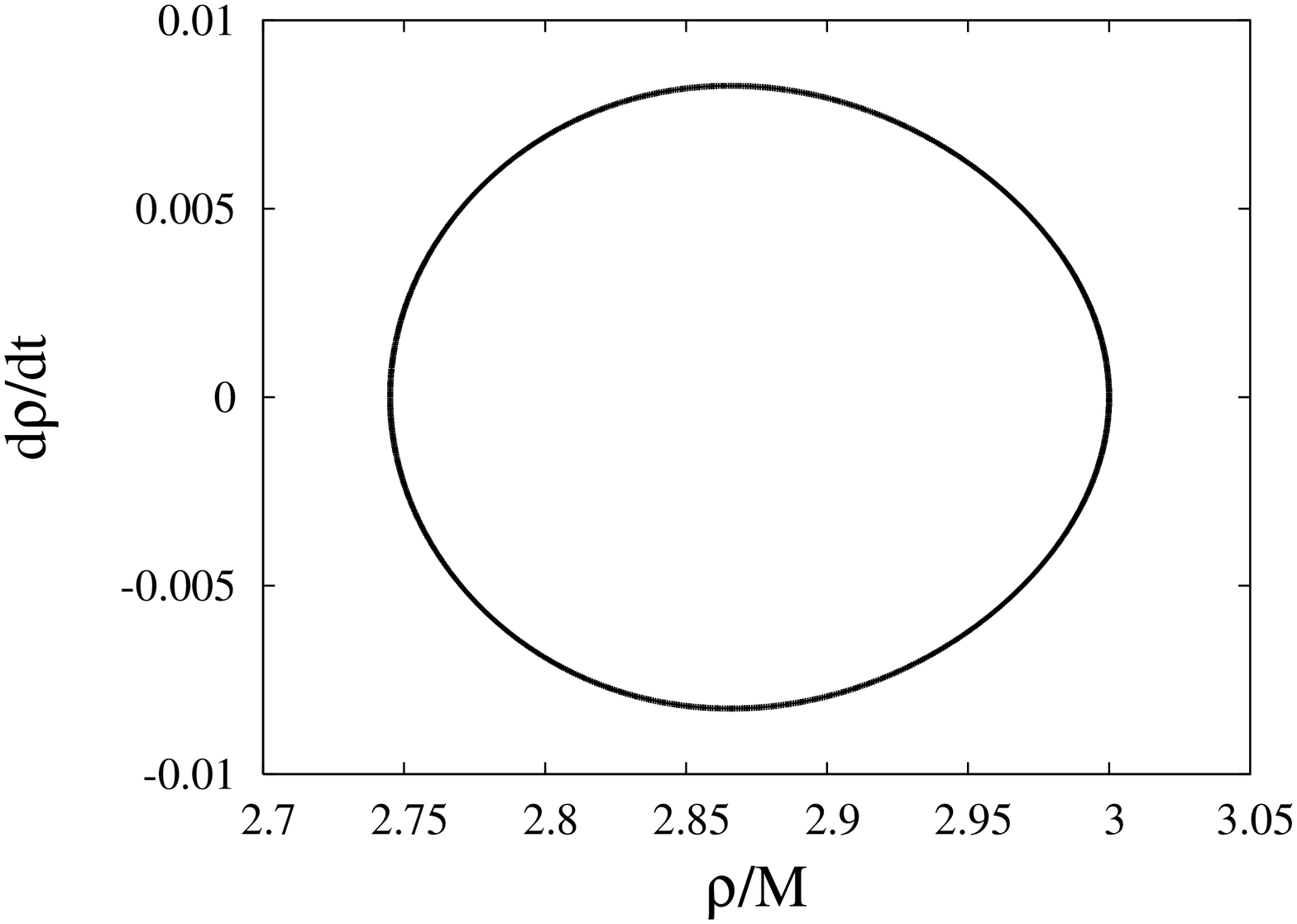} &
\includegraphics[keepaspectratio=true,height=3.5in,angle=-90]{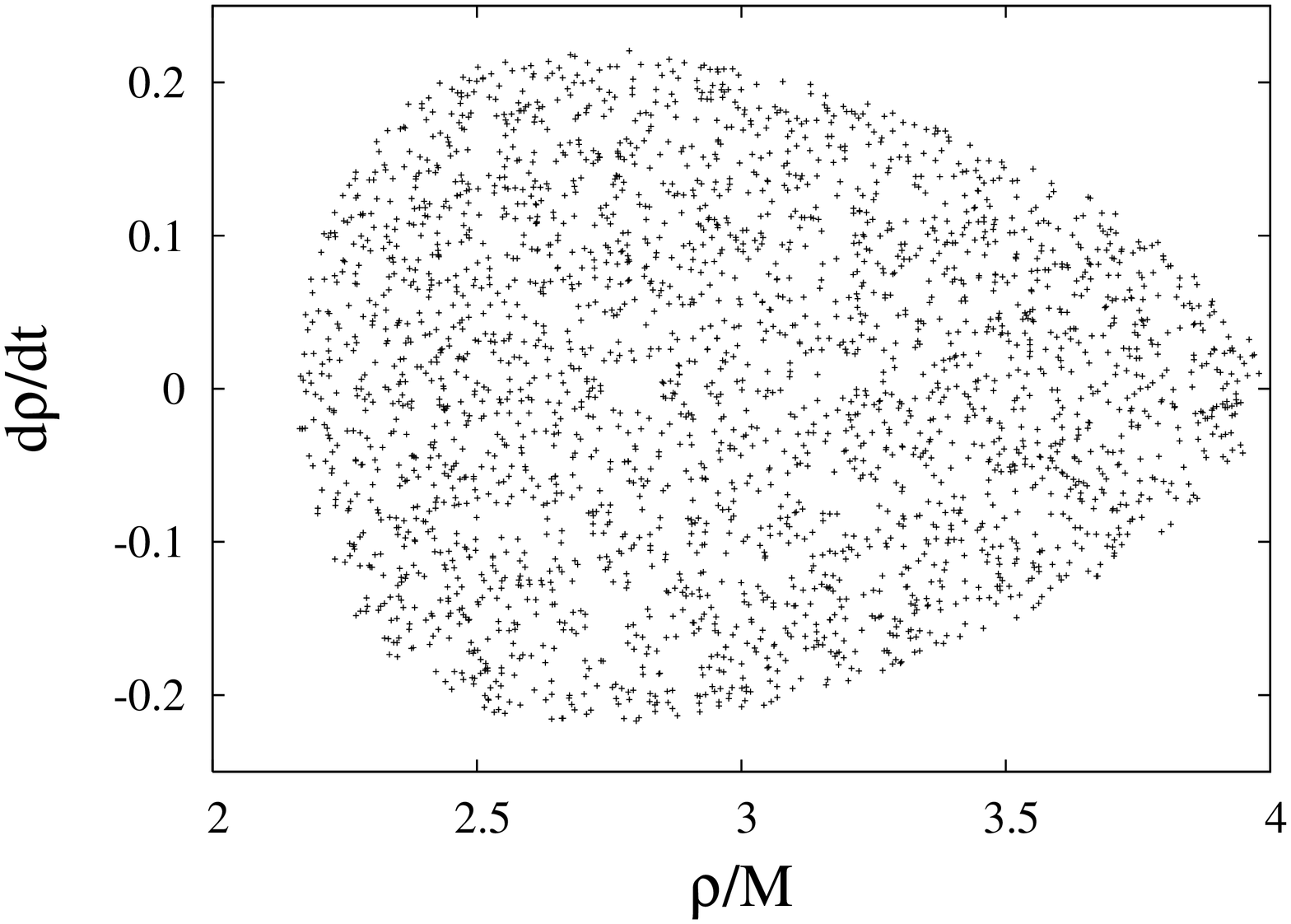} \\
\end{tabular}
\caption{Example of onset of chaos in the Newtonian quadrupole-octupole potential~\erf{quadoct}. All plots are for orbits which start with $\dot{\rho}=0=z$, $\rho/M=3$ and have specific angular momentum $L_z = 1.7M$. The left hand panels are for energy $E=0.82$, while the right hand panels have energy $E=0.81$. The top two plots show zeros of the effective potential, $V_{\rm eff}=0$, as defined by equation~\erf{quadoctveff}, and the paths followed by the orbits in the $(\rho, z)$ plane. The bottom two plots are Poincar\'{e} maps for crossings of the $z=0$ plane in each case.}
\label{quadoctfig}
\end{figure}

The classic example from astrophysics of a system that exhibits chaos in classical (Newtonian) gravity is the two dimensional H\'{e}non-Heiles potential $V(r,\theta) = r^2/2 + r^3 \sin(3\theta)/3$ (see~\cite{contop} for example). Gu\'{e}ron and Letelier~\cite{GL01} also found chaos in the Paczy\'{n}ki-Witta potential ($\Phi = M/(r-r_S)$, where $r_S = 2M$ is the Schwarzschild radius) with a dipolar perturbation. Neither of these spacetimes is reflection symmetric, so for a better analogy to the relativistic spacetimes considered in this paper, we examine the Newtonian quadrupole-octupole potential
\be
\Phi(\rho,z) = -\frac{M}{r} - \frac{M_2}{2\,r^3} \left( 1- 3\frac{z^2}{r^2} \right) + \frac{M_4}{8\,r^5}\left(35 \frac{z^4}{r^4} - 30 \frac{z^2}{r^2} + 3 \right) .
\label{quadoct}
\ee
Here, $M$, $M_2$ and $M_4$ denote the monopole (mass), quadrupole and octupole multipole moments of the potential. Stationarity and axisymmetry ensure that energy $E$ and angular momentum $L_z = r^2 \rmd\phi/\rmd t$ are conserved as usual, which leads us to the Newtonian analogue of the effective potential equation~\erf{Veff}
\be
\frac{1}{2}\left( \left(\frac{\rmd r}{\rmd t}\right)^2 +  \left(\frac{\rmd z}{\rmd t}\right)^2\right) = V_{\rm eff}(E, L_z, \rho, z) = \frac{1}{2}\left(E^2 - 1\right) -\frac{L_z^2}{2\rho^2} -\Phi(\rho,z)
\label{quadoctveff}
\ee
where we have replaced the standard Newtonian energy by the relativistic expression $(E^2 - 1)/2$ for consistency with~\erf{Veff}. The equation of motion in this potential takes the usual form $\rmd^2 {\bf r}/\rmd t^2 = -\nabla \Phi$. If we take the multipole moments to have the values $M_2 = 2M^3$ and $M_4 = 10M^5$, and choose the angular momentum to be $L_z=1.7M$, we find that for a range of values of the energy $E$, bound orbits occur quite close to the origin. For sufficiently large values of $E$, there is a single allowed region for motion (defined by $V_{\rm eff} \ge 0$). Orbits in that regime appear to be regular, and show closed Poincar\'{e} maps. If the energy is reduced, the allowed region eventually splits into two separate regions, one bounded away from $r=0$, and one connected to $r=0$. Orbits in the outermost region after this transition exhibit ergodic behavior. In Figure~\ref{quadoctfig} we show four plots. Two of these plots are for an orbit with $E=0.82$, which exhibit regular behavior. The other two are for $E=0.81$ and exhibit ergodic behavior. We choose the initial conditions of both orbits to be $\dot{\rho} = 0 =z$ and $\rho = 3 M$, with $\dot{z}$ determined from the assigned energy~\erf{quadoctveff}. The upper panels in the figure show the orbit in the $(\rho, z)$ plane, and the boundary of the allowed region of motion (defined by $V_{\rm eff}=0$). The lower panels show Poincar\'{e} maps for the two orbits. The ergodicity of the orbit with $E=0.81$ is quite evident from the Poincar\'{e} map. We also find that this orbit fills up the entire allowed range of $\rho$ and $z$. By contrast, the regular orbit with $E=0.82$ explores only a narrow torus in space.

A thorough examination of when ergodicity appears in this potential, as a function of energy, angular momentum and the multipole moments $M_2$ and $M_4$ is peripheral to the focus of this paper. However, the results presented here provide a Newtonian example to which we can compare the relativistic results of Section~\ref{integrals}.

\section{Weak Field Precessions}
\label{wfprec}
\subsection{Relativistic Precession}
In Boyer-Lindquist coordinates, the energy, angular momentum and rest-mass conservation equations~\erf{genconsts}--\erf{masscons} for geodesic motion in the Kerr metric can be used to derive the equation of motion in the form (see for instance~\cite{chandra83})
\ba
\frac{1}{2}\left(\left(\frac{\rmd r}{\rmd t}\right)^2 + \Delta  \left(\frac{\rmd \theta}{\rmd t}\right)^2\right) &=& \nonumber \\ && \hspace{-1.75in}\frac{\left(E(r^2+a^2)-aL_z\right)^2-\Delta\left(r^2+(L_z-aE)^2+L_z^2\cos^2\theta+a^2\cos^2\theta(1-E^2)\right)}{2\left( E((r^2+a^2)^2/\Delta - a^2\sin^2\theta) -2MaL_zr/\Delta\right)^2} \label{Kerreffpot}
\ea
where $\Delta = r^2 - 2Mr+a^2$, and $a = M\chi$. The prograde equatorial circular orbit at radius $r$ has energy and angular momentum
\ba
E &=& \frac{1-2v^2+av^3/M}{\sqrt{1-3v^2+2av^3/M}} \\
L_z &=&rv \frac{1-2av^3/M+a^2v^4/M^2}{\sqrt{1-3v^2+2av^3/M}} 
\ea
where $v^2 = M/r$. The frequency of a prograde circular orbit is given by
\be
\Omega_\phi= \frac{\rmd\phi}{\rmd t} = \frac{\sqrt{M}}{r^{3/2} + a\sqrt{M}}.
\ee
The epicyclic frequencies for radial and vertical perturbations of the orbit are given by the second derivatives of the right hand side of equation~\erf{Kerreffpot} with respect to $r$ and $\theta$ (the right hand side of Eq.~\erf{Kerreffpot} is the effective potential for the Kerr spacetime). To obtain the form of these frequencies in the weak field, we wish to expand in $1/r$. With some manipulation and keeping terms up to $r^{-5}$ only, we obtain the expansion
\ba
\Omega_\rho^2 &=& \frac{M}{r^3} - 6\frac{M^2}{r^4}+6\chi\frac{M^{5/2}}{r^{9/2}}-3\chi^2\frac{M^3}{r^5} + \cdots \\
\Omega_z^2 &=& \frac{M}{r^3}-6\chi\frac{M^{5/2}}{r^{9/2}}+3\chi^2\frac{M^3}{r^5} + \cdots
\ea
where we use $\Omega_\rho$, $\Omega_z$ to denote the radial and vertical epicyclic frequencies to be consistent with the results earlier in the paper. With further manipulation, expressions for the precessions, $p_X$, as a function of the orbital frequency, $\Omega_\phi$, may be derived
\ba
p_{\rho} &=& 3 \left(M\Omega_\phi\right)^{\frac{2}{3}} -4\chi \left(M\Omega_\phi\right)+\frac{3}{2}\left(9+\chi^2\right)  \left(M\Omega_\phi\right)^{\frac{4}{3}} -34\chi \left(M\Omega_\phi\right)^{\frac{5}{3}} \nonumber \\ && +\frac{1}{2}\left(135+67\chi^2\right)  \left(M\Omega_\phi\right)^{2}  + \cdots \\
p_{z} &=& 2\chi  \left(M\Omega_\phi\right) - \frac{3}{2}\chi^2 \left(M\Omega_\phi\right)^{\frac{4}{3}} +8 \chi^2  \left(M\Omega_\phi\right)^{2} +\cdots \label{Kerrprec}
\ea
Results for retrograde orbits may be obtained by the substitutions $\chi \rightarrow -\chi$, $\Omega_\phi \rightarrow -\Omega_\phi$ and $L_z \rightarrow -L_z$ in the above expressions (NB $\Omega_{\phi} < 0$ for retrograde orbits, so $-\Omega_{\phi}$ is equivalent to $|\Omega_{\phi}|$).

\subsection{Precession due to a Quadrupole Moment}
The precession induced by a quadrupole moment can be derived using the Newtonian quadrupole potential
\begin{equation}
\Phi = -\frac{M}{r} -\frac{1}{2} \frac{Q}{r^3} \left(1-3\frac{z^2}{r^2}\right) .
\end{equation}
Here $r = \sqrt{x^2+y^2+z^2}$ is the distance from the origin, $z$ is the vertical coordinate and we will use $\rho = \sqrt{x^2+y^2}$ to denote the cylindrical polar radial coordinate.
The radial equation of motion in this potential takes the form
\be
\frac{1}{2} \left( \frac{\rmd \rho}{\rmd t} \right)^2 = E-\frac{L_z^2}{2\rho^2} + \frac{M}{r} +\frac{1}{2} \frac{Q}{r^3} \left(1-3\frac{z^2}{r^2}\right) \label{newtquadeffpot}
\ee
and the energy, angular momentum and orbital frequency of a circular, equatorial orbit with radius $\rho$ are
\ba
E &=& \frac{Q}{4\rho^3} - \frac{M}{2\rho}\nonumber \\
L_z &=& \sqrt{M\rho + \frac{3}{2} \frac{Q}{\rho}}\nonumber \\
\Omega_{\phi} &=& \sqrt{\frac{M}{\rho^3} + \frac{3}{2} \frac{Q}{\rho^5}}
\ea
Differentiating Eq.~\erf{newtquadeffpot} twice with respect to $\rho$ and $z$, we find the epicyclic frequencies take the form
\ba
\Omega_\rho^2 &=& \frac{M}{\rho^3} -\frac{3}{2}Q\frac{M}{r^5} + \cdots \\
\Omega_z^2 &=& \frac{M}{\rho^3}+\frac{3}{2}Q\frac{M}{r^5} + \cdots .
\ea
Hence we derive the precession frequencies
\ba
p_{\rho} &=& -\frac{3}{2} \frac{Q}{M^3} \left(M\Omega_\phi\right)^{\frac{4}{3}}  +\cdots \nonumber \\
p_{z} &=& \frac{3}{2} \frac{Q}{M^3} \left(M\Omega_\phi\right)^{\frac{4}{3}} +\cdots .
\ea
The lowest order form of these expressions was also given in Collins and Hughes~\cite{CH2004}, although they expressed the precession in terms of a radial coordinate, rather than the observable $\Omega_{\phi}$. We also use a slightly different definition for the quadrupole moment $Q$ so that it is consistent with $Q=-\chi^2 M^3$ for the Kerr metric. As we would expect, the leading-order terms in these expressions agree with the leading order terms in $\chi^2$ in the Kerr expressions. 

Combining this result with Eq.~\erf{Kerrprec}, we obtain the weak-field precessions for the Manko-Novikov solution with spin parameter $\chi$ and excess quadrupole moment $q$
\ba
p_{\rho} &=& 3 \left(M\Omega_\phi\right)^{\frac{2}{3}} -4\chi \left(M\Omega_\phi\right)+\frac{3}{2}\left(9+\chi^2 + q\right)  \left(M\Omega_\phi\right)^{\frac{4}{3}} -34\chi \left(M\Omega_\phi\right)^{\frac{5}{3}} \nonumber \\ && +\frac{1}{2}\left(135+67\chi^2 +39q\right)  \left(M\Omega_\phi\right)^{2}  + \cdots \nonumber \\
p_{z} &=& 2\chi  \left(M\Omega_\phi\right) - \frac{3}{2}\left(\chi^2 +q\right)\left(M\Omega_\phi\right)^{\frac{4}{3}} +\left(8 \chi^2 -3q\right)  \left(M\Omega_\phi\right)^{2} +\cdots \label{MNweakprec} .
\ea
In the above, the lowest order term that is omitted is the order at which the excess current quadrupole moment would first contribute. This result is also given in Ryan~\cite{Ryan1995}, although he quotes an expression for $\tilde{\Omega}_\rho/\Omega_{\phi}$, where $\tilde{\Omega}_\rho$ is equal to $\Omega_{\phi} - \Omega_\rho$. Our result is consistent with his once this is taken into account. We note that some of the terms in expression~\erf{MNweakprec} come from relativistic corrections to the effect of the quadrupole moment. These cannot be derived using only the results quoted in this appendix, but are given in Ryan's paper~\cite{Ryan1995}.


\end{document}